\pdfoutput=1
\documentclass{JINST}
\usepackage[pdftex]{graphics}
\usepackage{subfig}
\usepackage{wrapfig}

\title{\huge{ \textbf{The Trigger and Timing System of the Double Chooz Experiment}}}

\author{F. Beissel$^a$$^{\dagger}$, A. Cabrera$^b$, A. Cucoanes$^a$$^c$, J.V. Dawson$^b$, D. Kryn$^b$, C. Kuhnt$^a$, S. Lucht$^a$\thanks{Corresponding author.\newline$^\dagger$ deceased on 16.6.2012.\newline$^c$ now at: SUBATECH, CNRS/IN2P3, Universit\`{e} de Nantes, F-44307 Nantes, France.\newline$^d$ now at: Max-Planck-Institut f\"{u}r Kernphysik, 69029 Heidelberg, Germany.}~, B. Reinhold$^a$$^d$, M. Rosenthal$^a$, S. Roth$^a$, A. Stahl$^a$, A. St\"{u}ken$^a$, C. Wiebusch$^a$\\
\llap{$^a$} III. Physikalisches Institut, RWTH Aachen University\\
  52056 Aachen, Germany\\
  E-mail: \email{sebastian.lucht@physik.rwth-aachen.de}\\
\llap{$^b$} APC, AstroParticule et Cosmologie, Universit\'{e} Paris Diderot, CNRS/IN2P3\\

}

\abstract{\noindent Modern precision neutrino experiments like Double Chooz require a highly efficient trigger system in order to reduce systematic uncertainties. The trigger and timing system of the Double Chooz experiment was designed according to this goal.\\
\noindent The Double Chooz trigger system is driven by the basic idea of triggering on multiple thresholds according to the total visible energy and additionally triggering on the number of active photomultiplier tubes (PMTs) in the detector. To do so, the trigger system continuously monitors the analogue signals from all PMTs in the detector. The amplitudes of these PMT-signals are summed for groups of certain PMTs (group signals) and for all PMTs (sum signal), respectively. The group signals are discriminated by two thresholds for each input channel and four thresholds for the sum signal. The resulting signals are processed by the trigger logic unit which is implemented in a FPGA. In addition to the proper trigger, the trigger system provides a common clock signal for all subsequent data acquisition systems to guarantee a synchronous readout of the Double Chooz detectors. The present design of the system provides a high flexibility for the applied logic and settings, making it useful for experiments other than Double Chooz.\\
\noindent The Double Chooz trigger and timing system was installed and commissioned in 2011. This article describes the hardware of the trigger and timing system. Furthermore the setup, implemented trigger logic and performance of the trigger and timing system for the Double Chooz experiment is presented.\\

}


\dedicated{Dedicated to Franz Beissel and his family.}

\begin{document}

\section{Introduction}

Double Chooz \cite{DCProposal} is a reactor anti-neutrino experiment located next to the nuclear power plant at Chooz, France. Its main physics goal is the measurement of the neutrino mixing angle $\theta_{13}$. The main challenge for the experiment, as for all reactor anti-neutrino experiments, is the reduction of systematic uncertainties.\\
\noindent Double Chooz uses the inverse beta decay for the detection of the anti-neutrinos comming from the reactor cores.

\begin{equation}
\label{IBD}
\overline{\nu_e}\,+\,p\,\longrightarrow \,e^+\,+\,n
\end{equation}

\begin{figure}[htb]
\centering
\includegraphics[angle=0,width=0.9\textwidth]{{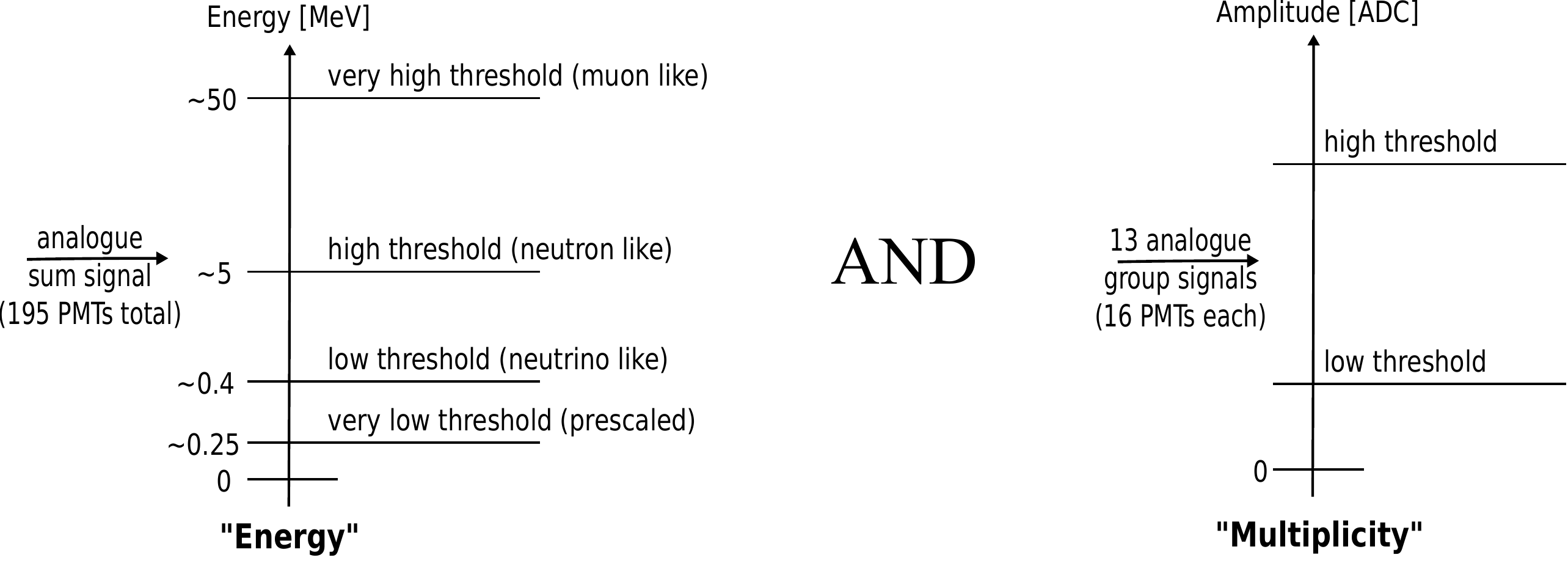}}
\caption{Schematic principle of the trigger decision of the Double Chooz experiment. The detector is triggered by a logical AND of the deposited energy inside the detector and a multiplicity condition representing the number of active groups of PMTs. The analogue sum signal is the summation of all group signals and its pulse height corresponds to the energy deposited in the detector.}
\label{E_Mult_Principle}
\end{figure}

\noindent An anti-neutrino interacting with a proton in the detector produces a positron with a continuous energy distribution from 1.02 to 12 MeV and a neutron.  The neutron is captured on Gd releasing $\gamma s$ with an average energy of 8 MeV \cite{DCFirstPub, DC2ndPub}. This happens after a characteristic time of typically $28\, \mu s$.\\
\noindent Main requirement for the rigorous reduction of systematic uncertanties is that the trigger system of the Double Chooz experiment has to trigger both the prompt and the delayed signal with a high efficiency ($> 99\%$) and that this trigger efficiency can be measured with sub-percent precision \cite{DCProposal}. Furthermore, the system has to be reliable, i.e. trigger failures must be rare and easily detectable.\\
\noindent For Double Chooz, the system creates a common clock signal for the whole detector and distributes a time stamp, unique event number and event classification for each triggered event. Also it is able to process the signals from external trigger sources such as calibration systems.\\
\noindent The trigger system is based on a redundant trigger decision. Input to the trigger are the analogue signals of the PMTs in the detector. The signals are proportional to the energy deposited in the detector. Each input signal itself is the sum of typically 16 PMTs (group signal) and it is discriminated by a low and a high threshold individually. The number of active groups is used for a multiplicity condition. The analogue sum of all group signals is discriminated by four different thresholds corresponding to a certain deposited energy inside the detector. The final trigger decision is formed by a logical AND of the deposited energy and a multiplicity condition on the active number of PMT groups (see Figure \ref{E_Mult_Principle}). Further, another level of redundancy is introduced by using two identical systems for the inner detector, each connected to half of the PMTs such both systems monitor the same detector volume.\\
\noindent The output status of the sum thresholds can be combined for a rough event classification at the trigger level which can be used for an online data reduction. This classification allows for instance to distinguish between the positron and neutron subsignal of the anti-neutrino signature (cf. Equation \ref{IBD}).\\
\noindent The implementation of this trigger in Double Chooz is based on a logical AND of a multiplicity condition and a discrimination of an analog sum signal of the PMTs. This is different from other compareable experiments. RENO uses a trigger system based on PMT hit multiplicity only \cite{Reno} while Daya Bay triggers on PMT multiplicity or an exceeded threshold on the sum signal proportional to deposited energy in the detector \cite{DayaBay}. Another example of a trigger based on multiplicity only is the low background experiment Borexino \cite{Borexino}.\\
The advantage of a trigger decision based on the deposited energy and a multiplicity condition is the introduction of redundancy and reliability to the system e.g. hardware failures are easily detectable. A similar argument applies for our decision to use two systems for the inner detector combined in a logical OR.

\section{The Double Chooz Detector and the Data Acquisition System}
\label{DAQandDetector}

This section will give a brief introduction of the Double Chooz detectors and the concept of the data acquisition system in which the trigger and timing system is embedded. For a more detailed description of the detector the reader is referred to the following documents \cite{DCProposal}, \cite{DCFirstPub}, \cite{DC2ndPub}, \cite{LiquidsPaper}.

\begin{figure}[tb]
\centering
\includegraphics[angle=0,width=0.4\textwidth]{{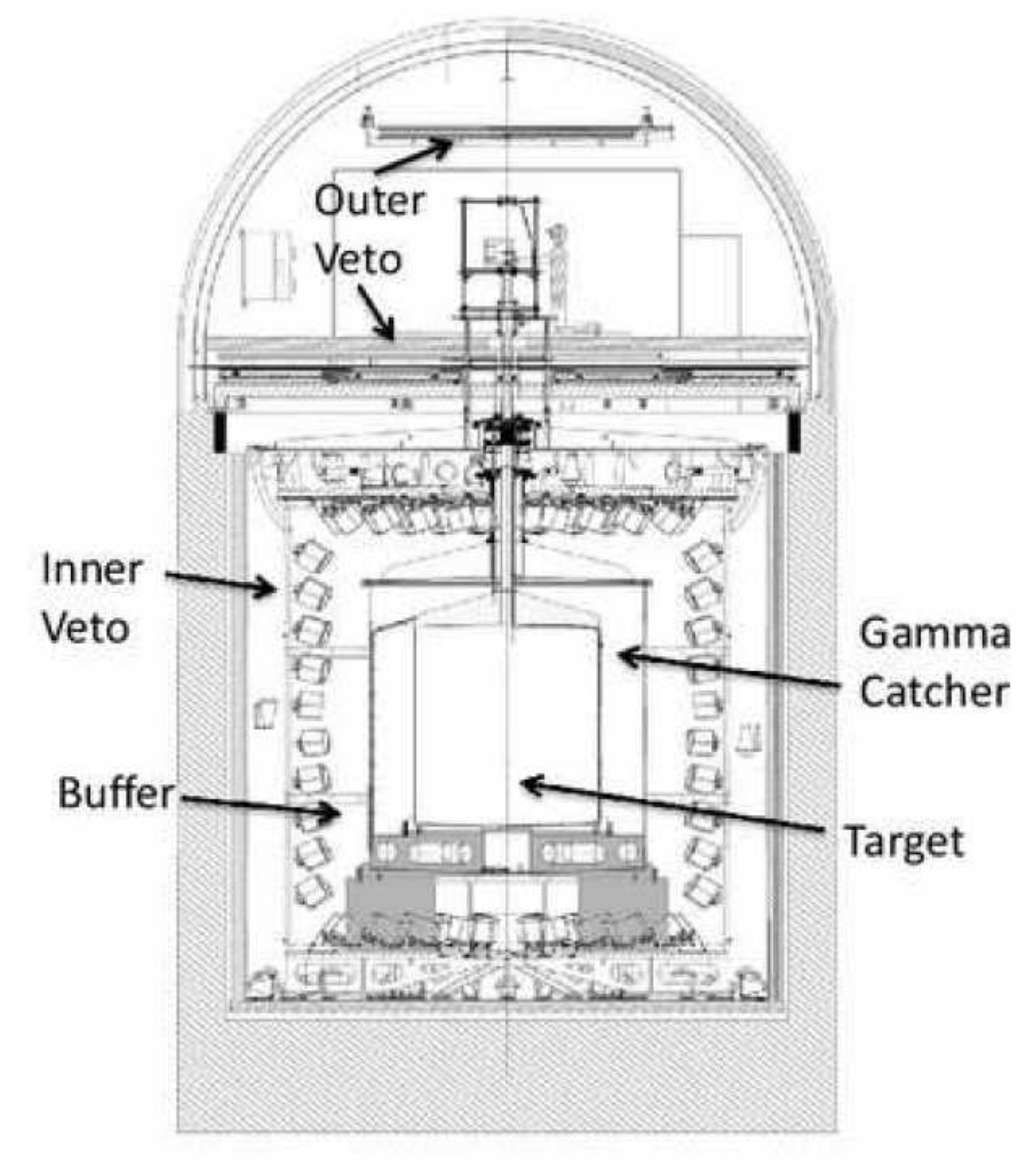}}
\caption{A cross-sectional view of the Double Chooz detector system \cite{DCFirstPub}. The inner detector volumes including the neutrino target, gamma catcher and buffer region are enclosed from the optically separated inner veto vessel. To reduce radioactive backgrounds from the surrounding rock the whole detector is surrounded by a steel shielding. Above the detector the outer veto, a muon tracking system, is shown.}
\label{Detector}
\end{figure}

\begin{figure}[htb]
\centering
\includegraphics[angle=0,width=0.9\textwidth]{{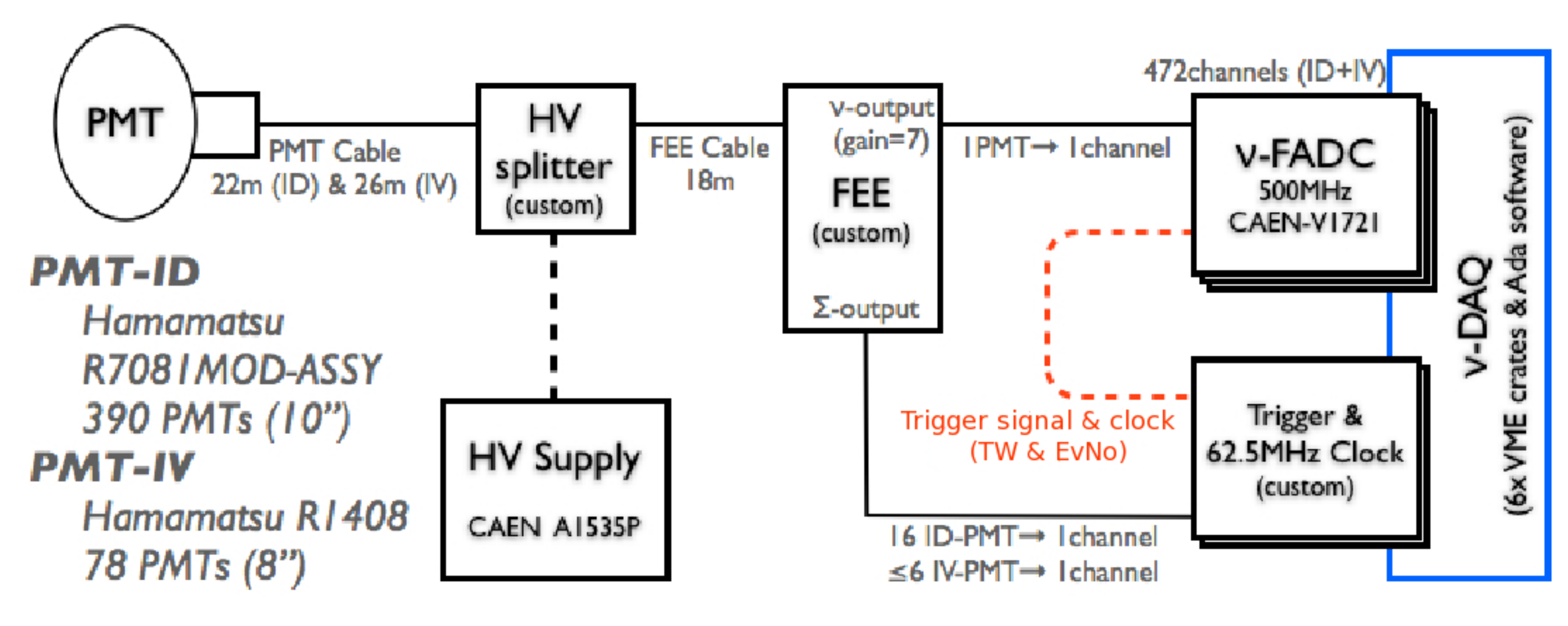}}
\caption[]{Schematic view of the data acquisition chain of the Double Chooz experiment \cite{DC2ndPub}. Interfaces to the outer veto are not shown.}
\label{DAQchain}
\end{figure}

\noindent Basically, the Double Chooz detector consists of four concentric cylinders, filled with different types of liquids. Figure \ref{Detector} shows the detector containing the inner veto, buffer, $\gamma$-catcher and neutrino target. The detector consists of two optically separated regions: The inner detector (ID) and the inner veto (IV). The ID contains the neutrino target, $\gamma$-catcher and buffer. The neutrino detection takes place inside the ID which is observed by 390 PMTs (Hamamatsu R7081MOD, 10-inch diameter \cite{PMTpaper}, \cite{japanese-PMTpub}). Enclosing the ID region, the IV detects cosmic muons and other backgrounds which enter the detector from outside. It is observed by 78 PMTs (Hamamatsu R1408, 8-inch diameter). The IV is surrounded by a 15 cm thick steel shielding to reduce radioactive background from the rock surrounding the detector.\\
A third subdetector system is the outer veto (OV), a muon tracking system. It consists of 44 plastic scintillator strips arranged in two layers. The first layer is installed directly above the inner veto vessel. The other one is installed under the ceiling of the experimental hall at a distance of approximately 5 m from the first layer.\\
\noindent The data acquisition records the waveforms of the PMTs with waveform digitizers ``FADC'' (\textbf{F}lash/ Fast \textbf{A}nalog to \textbf{D}igital \textbf{C}onverter) whenever a trigger signal is released by the trigger system. A schematic view of the data acquisition (DAQ) chain is shown in Figure \ref{DAQchain}.

\noindent All PMT signals are decoupled from the high voltage at the HV splitters. After decoupling, the PMT signals are transmitted to the Front-End Electronic (FEE) modules. Up to eight PMTs are connected to one FEE module. Each PMT signal is optimized (amplified, clipped, baseline restored and coherent noise filtered) by the FEE for digitization. The amplification factor of the $\nu$-FADC output is adapted to the dynamic range of the $\nu$-FADCs \cite{DissTA}. The $\nu$-FADCs are used to record neutrino like signals with a high resolution and therefore precise charge information.\\
Additionally, each FEE module provides an analogue sum of all connected PMTs. These signals are transmitted to the so-called ``stretcher circuit'' on the FEE modules. The circuit integrates the incoming charge of the PMTs over a time window of 100 ns. This time is adapted to the arrival time of photons which is determined by the detector geometry and the decay time of the scintillator as determined from dedicated MC simulation studies \cite{DissAC}. The resulting output signal from the ``stretcher circuit'' is the group signal. It is transmitted continuously to the trigger system. The amplitude of the group signal is proportional to the charge seen by the connected PMTs, collected within the time window of 100 ns. For the data published so far the waveforms over a time period of 256 ns are recorded whenever a trigger signal is released.\\ 
\noindent The trigger and timing system additionally distributes a common clock signal to all data acquisition components and distributes a time stamp, unique event number (EvNo) and a event classification (TW) for all triggered events (see Figure \ref{DAQchain}). The EvNo is used for an online validation of the synchronism between the $\nu$-FADCs and the trigger system. The TW allows an online data reduction based on the trigger decision (e.g. huge energy depositions which saturate the $\nu$-FADCs can be discarded).\\
The design of the trigger and timing system was done according to the following redundancy concept: The trigger decision is based on a logical AND of the discrimination of the analogue sum signal and a multiplicity condition of active PMT groups (see Figure \ref{E_Mult_Principle}) and a logical OR of two independent systems (here two Trigger Boards for the ID region).

\section{Hardware of the Trigger and Timing System}

This section will describe the hardware developed for the trigger and timing system in general. The design of the system provides a high flexibility for the applied logic and settings, making it useful for experiments other than Double Chooz. Parts of the system are used in AMANDA \cite{Laihem, Tepe_Thesis}, the \textit{Muon Track fast Tag} upgrade for the CMS detector \cite{Weinstock}, for the gas monitor chamber of the TPCs of the ND280 detector of T2K \cite{JS_Thesis} and in R\&D projects. A description of the setup for the Double Chooz experiment will be given in section \ref{DCsetup}.\\
\noindent The trigger and timing system is designed as a two level system. All components are VME (\textbf{V}ersa \textbf{M}odule \textbf{E}urocard) cards. The first stage is the so-called Trigger Board (TB) and the second the Trigger Master Board (TMB).\\
The signals from the detector are discriminated on the TB. The digital signals are processed by the logic (firmware) implemented in a ``FPGA'' (\textbf{F}ield \textbf{P}rogrammable \textbf{G}ate \textbf{A}rray, Xilinx Inc. model XC2V500 \cite{FPGARef}) on the TB and TMB. The FPGAs give a high flexibility. Modifications or additions of new features can be realized without any hardware modification. The current firmware of TB and TMB allow to modify various settings of the boards functionality via the VME-bus using a self customized software. Those changeable features are highlighted by a dotted line in the trigger logic schemes of both cards (see Figure \ref{TB_scheme} and Figure \ref{TMB_scheme}).
Each trigger module (TB and TMB) can use an external clock but is also equipped with an internal free running oscillator (XPRESSO model FXO-HC73 \cite{OscillatorRef}). Usually, the common system 62.5 MHz clock is distributed from the TMB to all TBs but it is also possible to run a TB or TMB standalone by using the internal oscillator.\\
The next sections will describe the single components and their functionality in detail.
\newpage
\subsection{Trigger Board}
\label{TB}

\begin{floatingfigure}[hl]
\includegraphics[angle=270,width=3.0cm]{{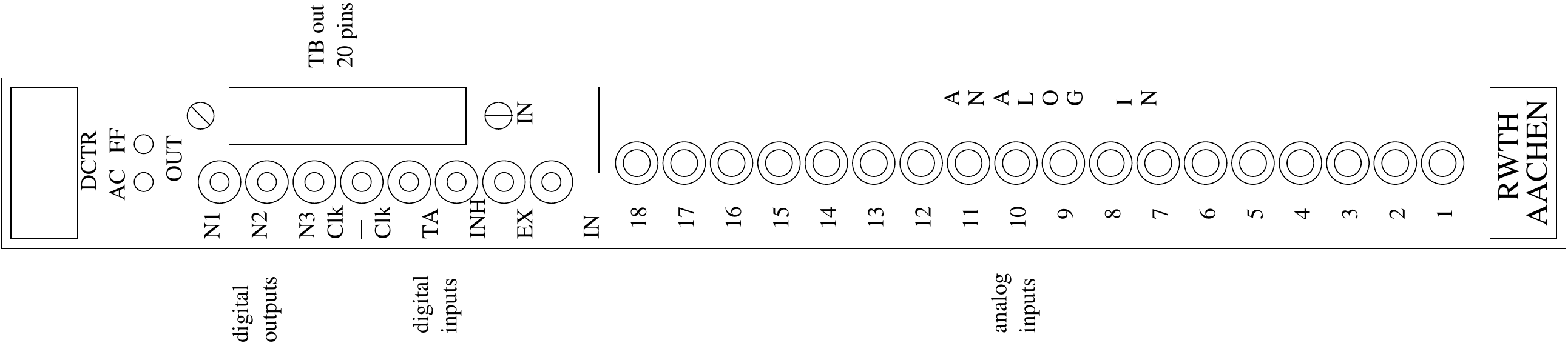}}
\caption{Trigger Board front panel.}
\label{TBfront}
\end{floatingfigure}

\begin{figure}[!tb]
\centering
\includegraphics[angle=0,width=1.\textwidth]{{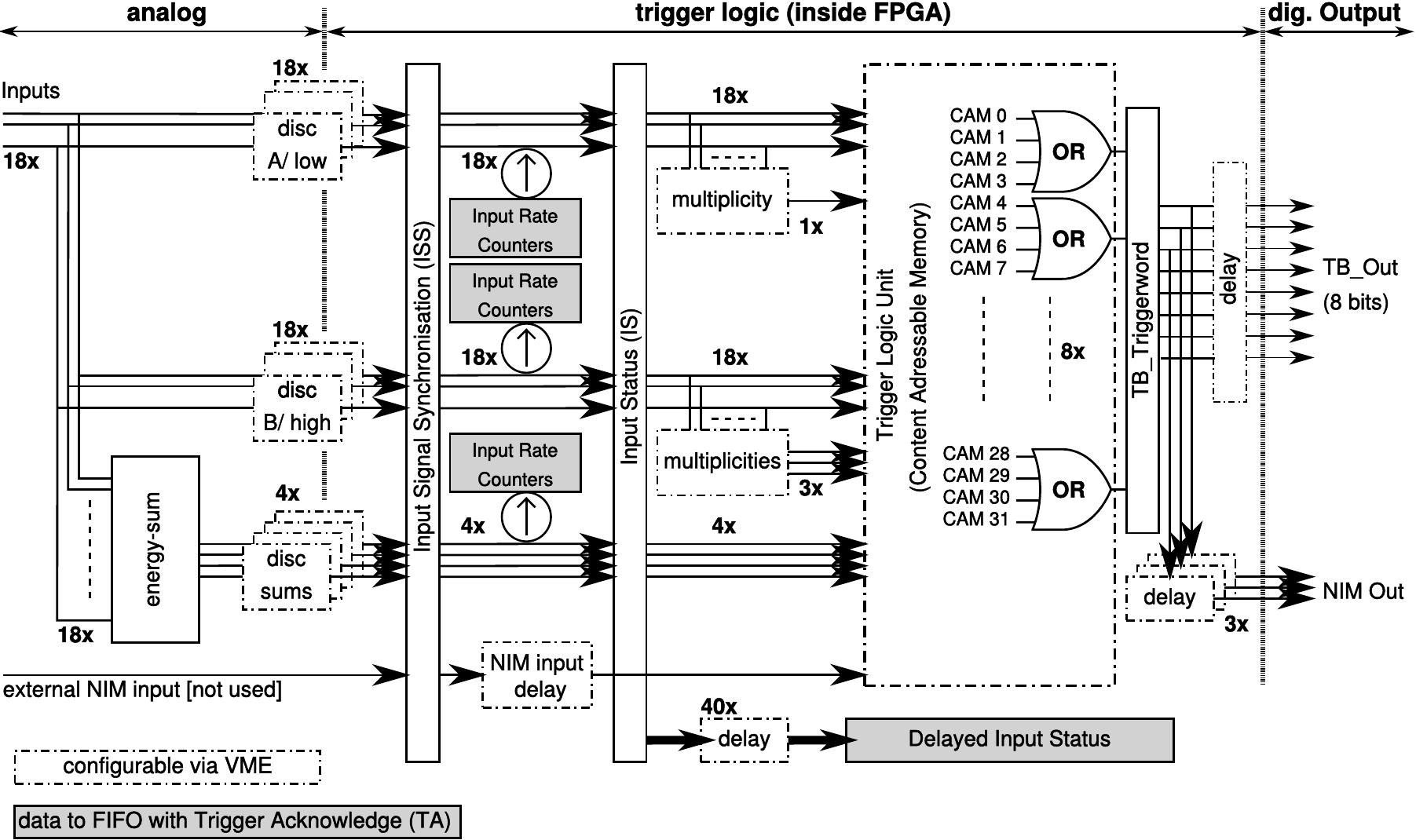}}
\caption{Scheme of the signal routing inside the TB (analogue part) and the TB firmware (digital part) implemented in the FPGA. Grey areas show information which is stored with each trigger signal. Dotted squares show free programmable parts or settings of the TB.}
\label{TB_scheme}
\end{figure}

\noindent The TB can be separated into an analogue and a digital part (see Figure \ref{TB_scheme}). The analogue part consists of various amplifiers and discriminators which convert the analogue input signals into digital signals for the trigger logic. The front panel of the TB is shown in Figure \ref{TBfront}.\\
\noindent There are 18 analogue input connectors (MCX, male) and four digital input connectors: One 2-pin LEMO connector socket (LVDS) for an external clock signal Clk and three LEMO connector sockets (Camac 00 series, NIM) for external signals (TA, INH and EX). The EX input can be used for an external trigger. The TA input receives the main trigger signal from the TMB. When it is active the TB data will be stored in its ``first in first out'' (FIFO) memory inside the FPGA. The INH receives the ``inhibit signal'' which is used for a synchronized start of the system. When it is active the TB is disabled. The main output is a 20-pin connector socket (LVDS). Usually this output includes the status information of the discriminators but the exact definition can be modified by the user (see Figure \ref{TB_scheme}). From the connector socket the output signal proceeds to the TMB by a flat ribbon cable. The lowest three bits of this output are transmitted in parallel via three LEMO connectors N1,..,N3 (NIM). The clock signal from the oscillator on the TB can be gripped from the 2-pin LEMO connector (LVDS) labled Clk.\\
The LED, labled ``AC'', indicates activity on the VME-bus. The ``FF'' LED gives a warning when the FIFO of the TB is full.\\
The signal logic inside the TB is schematically shown in Figure \ref{TB_scheme}.

\subsubsection{Analogue Part and Input Signal Synchronization}
\label{ANALOGUEandISS}
\begin{figure}[t]
\centering
  \includegraphics[angle=0,width=0.9\textwidth]{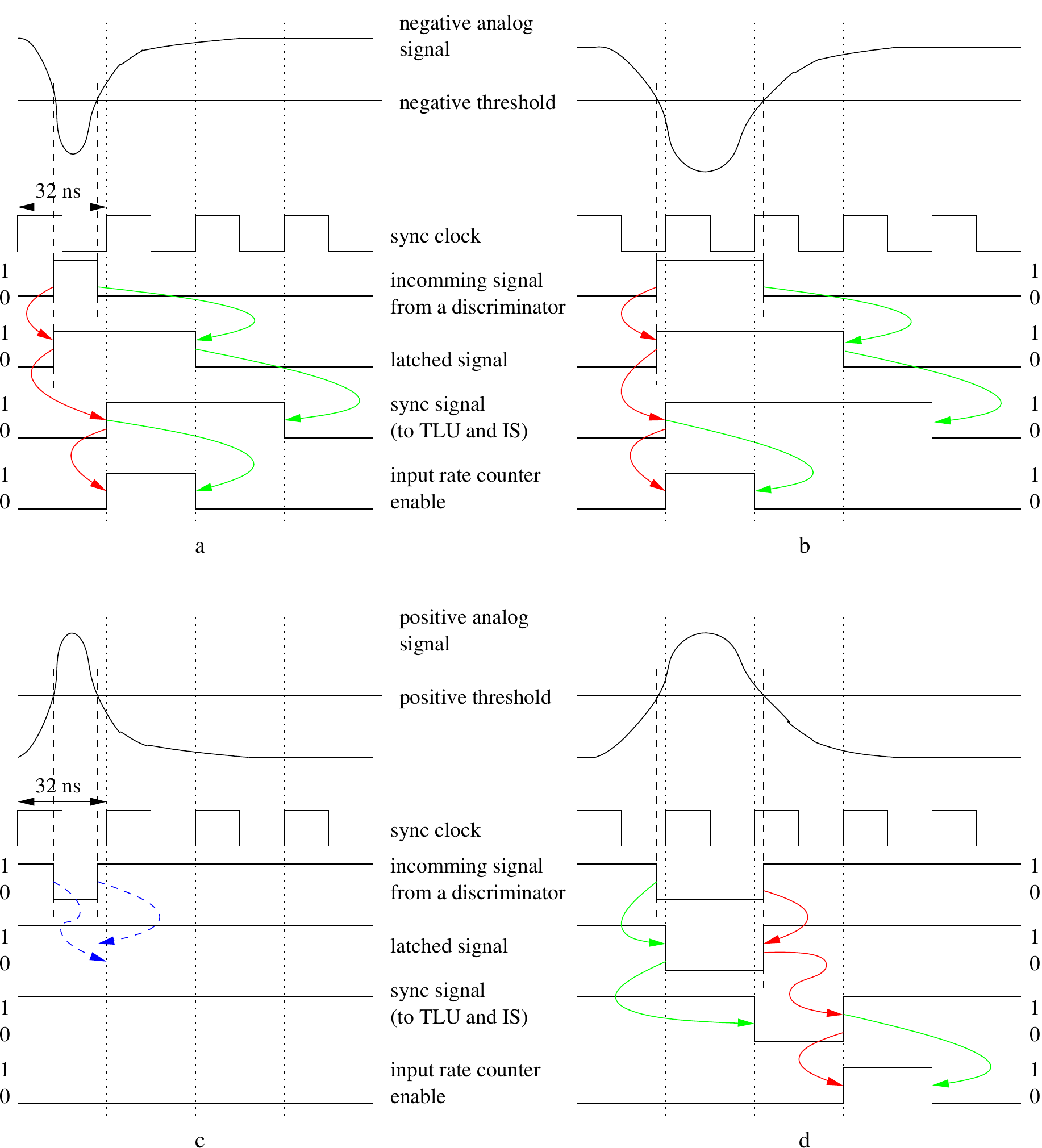}
  \caption{Functional principle of the ISS circuit. Figures a and b show the response to a negative input signal. Figures c and d show the response to a positive input signal. The pointers highlight which signal edges cause the corresponding next signals. Figure c shows an example for a positive input signal which will not be detected by the trigger system because the design is optimized for negative input signals.}
  \label{ISS}
\end{figure}

\noindent The analogue inputs are capacitively coupled (AC coupling) to an operational amplifier. The signals are discriminated twice by two discriminators (A and B) with a dynamic range of $\pm$1200 mV. A third signal is connected to summing amplifiers that produce the total sum of all input signals. The sum is discriminated by four discriminators (see Figure \ref{E_Mult_Principle}). Each discriminator of the sum operates in a different dynamic range due to different amplifiers on the signal line towards them. The thresholds of all discriminators are set through a twelve bit DAC (\textbf{D}igital \textbf{A}nalog \textbf{C}onverter). The discriminators generate a positive output while the threshold is exceeded. These signals are asynchronous with respect to the system clock. The input status synchronization (ISS) circuit inside the FPGA (see Figure \ref{ISS}) synchronizes them.
The ISS uses the sync clock, which is derived from the system clock, with a doubled clock cycle length (32 ns). The signals are processed by several flip-flops in the ISS circuit. First a latched signal is created. The latched signal is at least 32 ns long (see Figure \ref{ISS} a and b). It starts with the discriminator turning positive and ends with the next rising sync clock after the discriminator is back negative. At the next rising edge of the sync clock an active latched signal causes the begin of the corresponding sync signal. The sync signal ends with the clock cycle after the latched signal becomes inactive. This guarantees a signal length of the sync signal of at least 64 ns in order to compensate for transit time variations in the trigger logic and in the detector. Figure \ref{ISS} shows schematically how the ISS responds to different input signals.\\
The trigger system was optimized for negative input signals. An inherent characteristic to the design of the ISS is that the behavior of the trigger system for negative input signals differs from the response to positive input signals. When the thresholds are set to positive voltages the discriminator outputs (and the corresponding signals created in the ISS circuit) are active as long as the input signal amplitude is below that threshold (see Figure \ref{ISS} c and d). This leads e.g. to the asymmetrical spectrum shape around the baseline in the detector spectrum shown later in Figure \ref{ThresholdScanDC_ID}.

\subsubsection{Trigger Logic Unit}
\noindent From the ISS the sync signals are propagated to the Trigger Logic Unit TLU (see Figure \ref{TB_scheme}). AND and OR conditions (and their negations) can be applied to the incoming signals. In addition to the sync signals four multiplicity condition signals are sent to the TLU (one on the A discriminators and three different on the B discriminators). The TLU consits of 32 Content Addressable Memory (CAM) units which can be modified by software. In each CAM unit AND conditions (and its negation) between all the input signals of the TLU can be defined. All signals can be connected to each CAM in any order. A CAM generates a positive output if its condition is fulfilled. Groups of four CAMs are combined into a logical OR producing one bit of the 8 bit output of the TB.\\
The 8 bit output and the corresponding NIM outputs are delayable in steps of 16 ns via software.\\
In order to test the TB the discriminators can be set by software. It is also possible to disconnect all inputs from the ISS to test the internal functionality of the TLU.

\subsubsection{Trigger Board Data}
\label{TBdata}  
The FIFO in the FPGA can store up to 128 events. If the FIFO is full the TB is still fully functional but no further data can/will be stored. Whenever the TB receives a trigger acknowledge (TA) the following data is transmitted to the FIFO and can be read via the VME-bus.

\begin{itemize}

\item \textbf{Input status (IS):} The output of all discriminators are sent to a shift register (delay line). Via the VME-bus it can be defined which of these registers (16 ns each) should be written to the FIFO. Hence, it is possible to store the IS corresponding to the time the trigger condition was fulfilled. The correct setting of the timing depends on cable length between the boards and the transit time through the boards themselves\footnote{Modifications of the firmware and thereby changes of the digital signal path inside the FPGA can have an impact on the total transit time} and has to be adjusted every time changes to the system are made. The IS information requires 2 * 18 bit for the single channel discriminators and 4 bit for the sum discriminators.

\item \textbf{Input rate counter (IRC):} Dedicated 16 bit counter counts the switching rate of the discriminators between two consecutive TA signals. The IRC information requires 2 * 18 * 16 bit for the single channel discriminators and 4 * 16 bit for the sum discriminators.

\item \textbf{Time difference counter (TDC):} Counts the clock cycles between two consecutive TA signals in steps of the system clock. The TDC information requires 32 bit.

\item \textbf{Event number (EvNo):} this counter is stored inside the FIFO and incremented with every TA signal. The EvNo information requires 32 bit.

\end{itemize}

\noindent When the TB is used in stand-alone mode (e.g. without a TMB) the TA signal can be internally derived from one of the NIM outputs or from a gate timer register. The gate timer can be set by the VME-bus to generate the TA internally in fixed time intervals.

\newpage

\subsection{Trigger Master Board}
\label{TMB}

\begin{floatingfigure}[!hl]
\includegraphics[angle=270,width=3.0cm]{{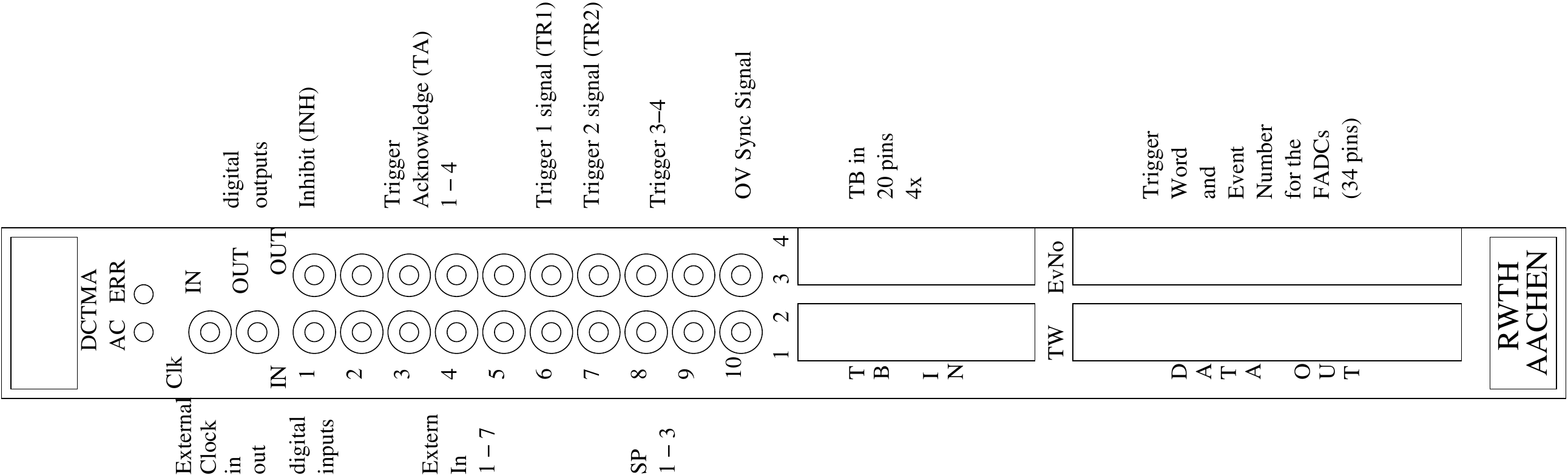}}
\caption{Trigger Master Board front panel.}
\label{TMBfront}
\end{floatingfigure}

\begin{figure}[tb]
\centering
\includegraphics[angle=0,width=1.\textwidth]{{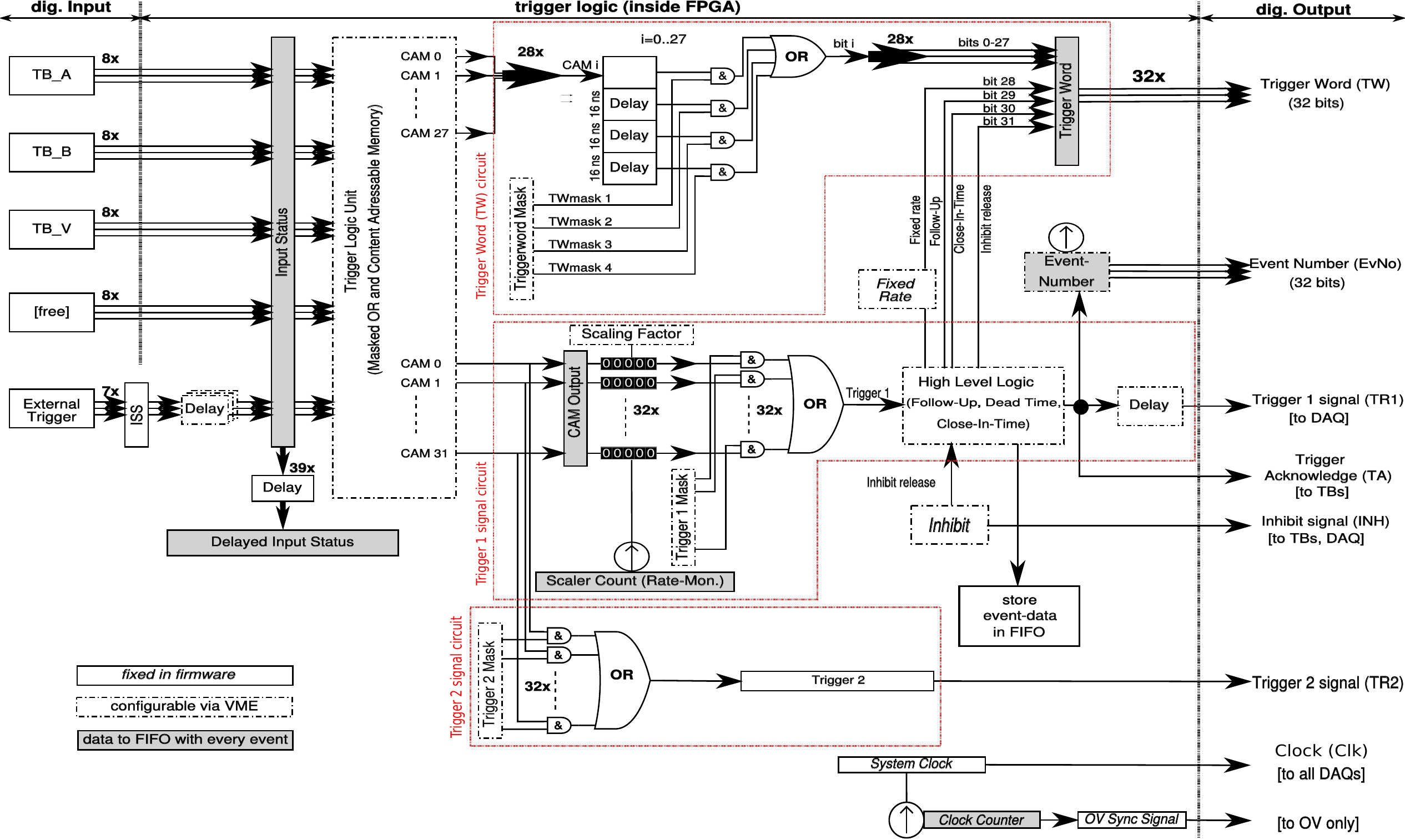}}
\caption{Scheme of the TMB firmware implemented in the FPGA}
\label{TMB_scheme}
\end{figure}

\noindent The TMB is the second stage of the trigger and timing system (see Figure \ref{TriggerSchematic}). It processes the incoming digital signals from up to four TBs and seven external trigger sources. On the front panel of the TMB are various connector sockets for digital input and output signals as shown in Figure \ref{TMBfront}: Two 2-pin LEMO connectors (LVDS) are for the output of the system clock signal (Clk) and the input of an external clock source, respectively. There are twenty LEMO connectors (Camac 00 series, NIM) for digital input and output signals. The digital output connectors provide (from top to bottom) the inhibit signal (INH), four Trigger Acknowledge (TA) output signals for up to four TBs and two outputs for the trigger signals (Trigger 1 and Trigger 2). The implemented logic for those trigger signal outputs will be described in section \ref{trig1} and \ref{trig2}. The remaining two sockets (Trigger 3-4) are disabled in the current firmware. The last socket is used for the so called OV sync signal which can be used for the synchronization of an external subsystem which is using its own clock. The ten LEMO connectors for digital input are reserved for external trigger signals. In the current firmware only the first seven input sockets (Extern In 1-7) are connected to the FPGA. The other sockets (SP 1-3) are disabled.\\
The output of connected TBs is transmitted to the TMB by two 20-pin connector sockets (LVDS) such that two TBs are connected to one socket. At the bottom of the front panel there are two 34-pin connector sockets (LVDS) providing the Trigger Word and Event Number which are described in section \ref{OutputsOfTS}. The LED labled ``AC'' , indicates activity on the VME-bus. The LED labled ``ERR'' gives a warning when the FIFO of the TMB is full.\\
A scheme of the logic implemented in the firmware and the signal path in the FPGA is shown in Figure \ref{TMB_scheme}. The incoming signals from external trigger sources are processed by an ISS circuit similar to the ISS of the TB (cf. section \ref{ANALOGUEandISS}). The sync clock on the TMB has a clock cycle length of 16 ns. All signals, external and from the TBs, are transmitted to the Masked OR. It creates the logical OR of all signals with a mask settable via the VME-bus. The Masked OR is followed by 32 CAMs. As on the TBs one can define logical AND (and its negation) conditions between all the input signals in any order for each CAM. Whenever a certain condition is fulfilled the CAM will generate an output signal.\\
The output from the CAMs is split and transmitted into three independent circuits.
 
\subsubsection{Trigger 1 Signal Circuit}
\label{trig1}

The outputs of all 32 CAMs are counted (Scaler Count, see Figure \ref{TMB_scheme}). This counter value is compared with the scaling factor which can be set through the VME-bus. If the counter reaches the scale factor an output is created and the counter reset. This allows to prescale a certain input to an acceptable rate for the DAQ (cf. section \ref{DClogic}). Afterwards the signals are compared with a software controlled 32 bit mask (Trigger 1 Mask). This mask acts as a switch for the CAM outputs. The unmasked signals are combined in a logical OR and transmitted to the high level logic (only for the trigger 1 signal), which provides additionally some special triggers:

\begin{itemize}
\item \textbf{Fixed rate trigger: } A random trigger with a fixed period from 16.384 $\mu$s up to 1.0738 s in steps of 16.384 $\mu$s.
\item \textbf{Follow up trigger: } The current logic prevents the system to create new triggers while a previous trigger condition is still fulfilled. By software it is possible to define a time between 32 ns and 528 ns (in steps of the system clock) after which a second trigger signal is fired if the previous condition is still fulfilled in order to avoid data loss and dead time.
\item \textbf{Close in time: } This function can prevent trigger signals which are very close to each other such that the corresponding data samples (e.g. from a waveform digitizer) would overlap. By software one can define a timewindow from 32 ns up to 528 ns after each trigger. Whenever a second trigger arrives in that time window it is delayed to the end of the time window. Instead of the close in time trigger one can also set a dead time from 16 ns up to 512 ns via the VME-bus. During this time no additional trigger can be created.
\item \textbf{Inhibit release: } Whenever the inhibit signal is released, a trigger will be generated. This function can be disabled by software.
\end{itemize}

\noindent These four special trigger signals have dedicated bits in the trigger word which is generated with each trigger signal (cf. section \ref{TW} and \ref{TMBdata}). The trigger 1 signal can be delayed up to 272 ns in steps of the system clock. With every generated trigger 1 signal all the TA outputs on the front plane are activated.\\
The CAM output signals, which have caused a trigger 1 signal, have to become inactive before a new trigger 1 signal can be created.

\subsubsection{Trigger 2 Signal Circuit}
\label{trig2}

With this circuit it is possible to define an additional trigger signal independently from the trigger 1 signal.\\
\noindent The 32 outputs of all CAMs are compared (logical AND) masked (see Figure \ref{TMB_scheme}, Trigger 2 Mask). All remaining signals are combined in a logical OR and transmitted to the trigger 2 signal output. To compensate the additional transit time of the trigger 1 signal caused by the high level logic (see above), the trigger 2 signal is delayed until both signals are synchronous.

\subsubsection{Trigger Word (TW) Circuit}
\label{TW}

The 32 bit TW provides information about a generated trigger 1 signal. It contains 4 bit with the information about the special triggers. The other 28 bits are reserved for the CAM output signals. The outputs of the first 28 CAMs of the TLU are transmitted to the trigger word circuit. Four shift registers store the CAM output of four consecutive clock cycles (with the rising edge of the clock signal the information is moved to the next shift register). For each shift register one can mask out certain bits. The masked content of the four shift registers is combined in a logical OR to build the trigger word (see Figure \ref{TMB_scheme}). It contains the information from 64 ns. The TW (and also the event number) is active at the corresponding output sockets 6 ns before the trigger 1 signal is released.

\subsubsection{Trigger Master Board Data}

The FIFO memory inside the FPGA can store up to 128 events. If the FIFO is full the TMB is still fully functional but no further data can/will be stored. Whenever the TMB generates a trigger 1 signal the data listed below is transmitted to the FIFO and can be read via the VME-bus.

\begin{itemize}
\label{TMBdata}
  
\item \textbf{Input status (IS):} The input status of all input lines of the TMB are stored in a shift register. A delay guarantees that the status of the clock cycle in which the trigger condition was fulfilled is stored with every event. The setting must be tuned with respect to the board's transit time in the respective firmware version (cf. section \ref{TBdata}). The external trigger signals can be delayed via software. The IS information requires 39 bit.

\item \textbf{CAM output:} The 32 bit output status of all CAMs.

\item \textbf{Scaler count:} Each CAM output is connected to a 16 bit counter. The counter is reset whenever its value matches the corresponding scaling factor (cf. section \ref{trig1}).

\item \textbf{Clock counter: } This 32 bit counter stores the number of clock cycles since the start of the data acquisition. When the counter reaches its maximum value it will restart counting from zero again.

\item \textbf{Event number (EvNo): } This counter is incremented with every generated trigger signal and requires 32 bit.

\item \textbf{Trigger word (TW:)} The 32 bit trigger word.

\end{itemize}

\section{The Setup for the Double Chooz Experiment}
\label{DCsetup}

\begin{figure}[htb]
\centering
\includegraphics[angle=270,width=0.9\textwidth]{{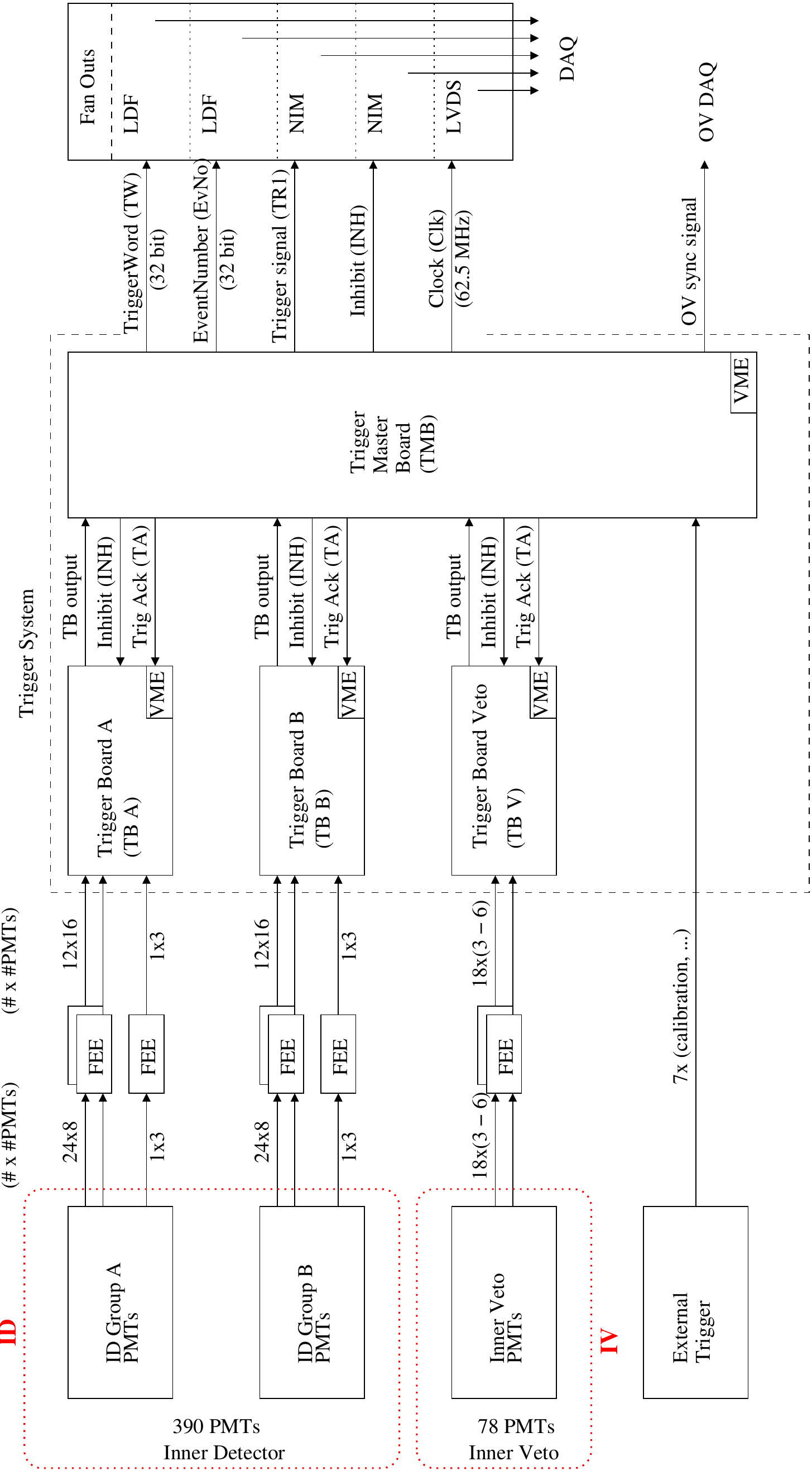}}
\caption[]{Schematic view of the trigger and timing system of the Double Chooz experiment. For the ID eight PMTs are connected to one FEE module. Two FEE modules together (16 PMTs) build one input signal of an ID Trigger Board (except for two FEE with only three connected PMTs). The ID PMTs are grouped such that both ID Trigger Boards observe the same detector volume introducing redundancy to the system. For the IV the PMTs are grouped such that each group represents a certain region of the IV vessel. The output of the Trigger Boards is then proceeded to the Trigger Master Board where the trigger decision is made. The output signals from the Trigger Master Board (cf. section \ref{OutputsOfTS}) are distributed to the DAQ by several fan-outs.}
\label{TriggerSchematic}
\end{figure}

\begin{figure}[htb]
\centering
\subfloat[]{
\includegraphics[angle=0,width=0.45\textwidth]{{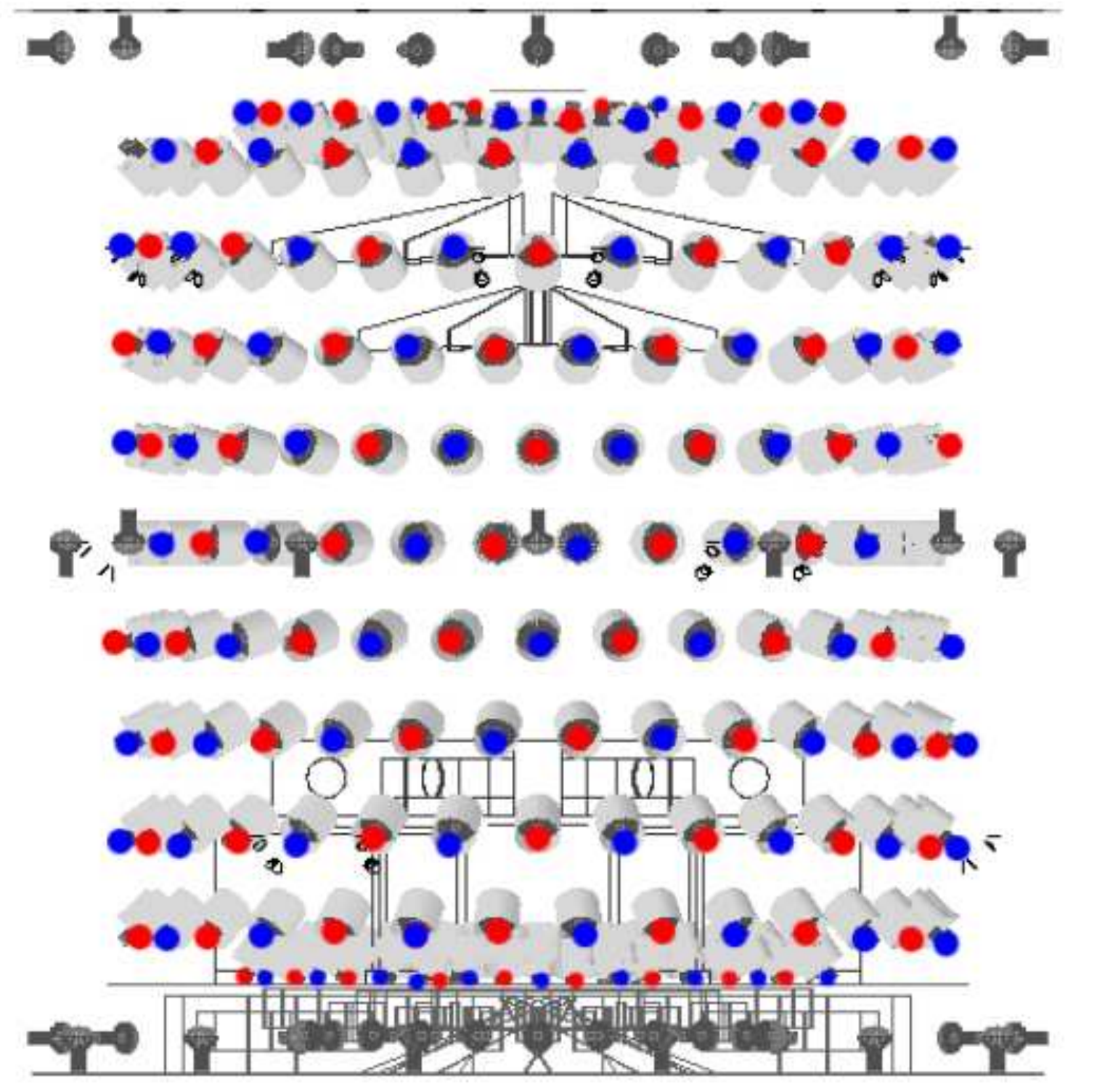}}
\label{IDgrouping}}
\subfloat[]{
\includegraphics[angle=0,width=0.45\textwidth]{{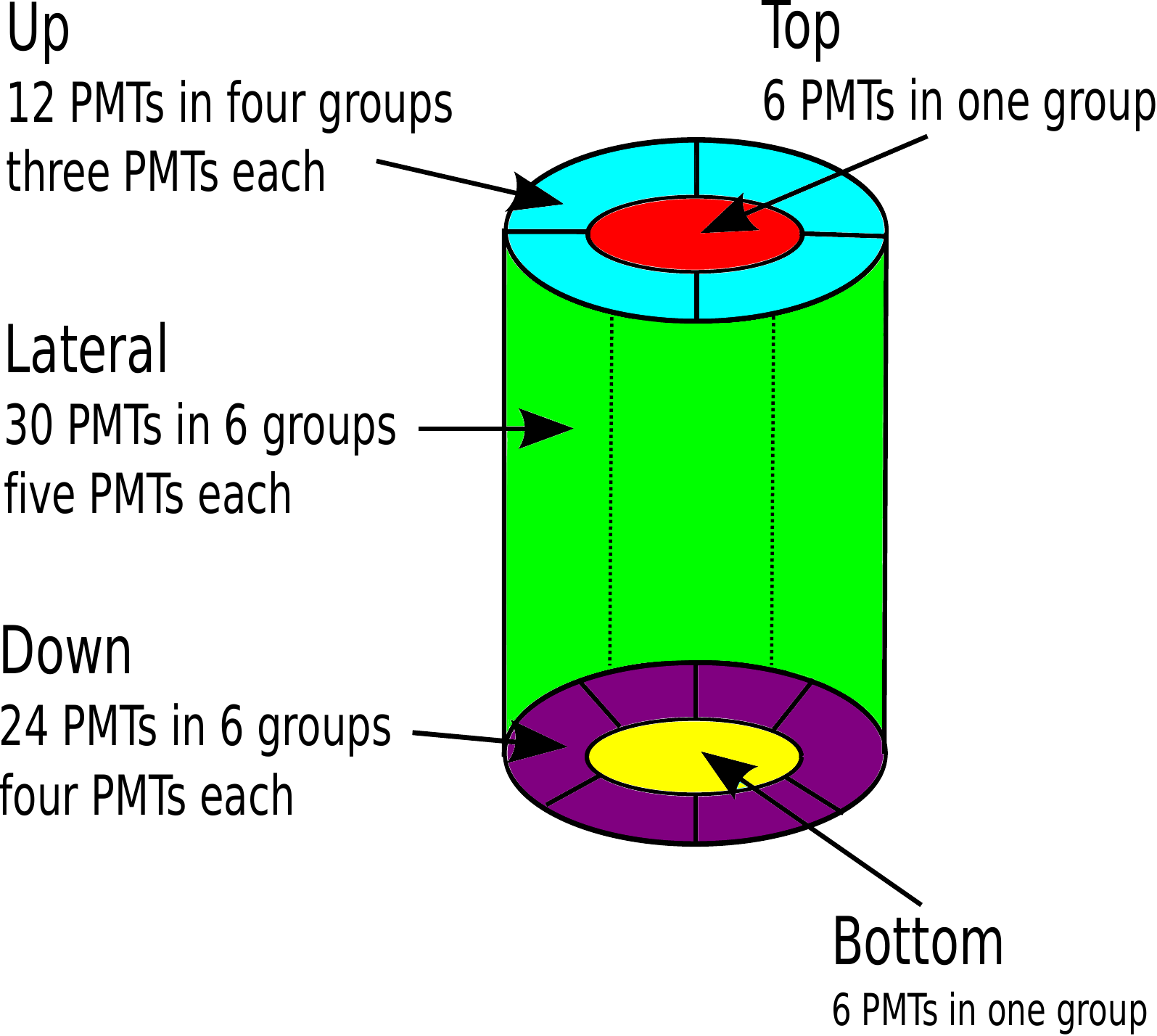}}
\label{IVgrouping}
}
\caption{(a): Grouping of the ID PMTs. Red (blue) PMTs are connected to the same TB. Both boards observe the same volume of the detector. (b): Schematic drawing of the PMT grouping for the IV. There are 18 different PMT groups, each of them monitoring a certain region of the IV. The colours show which groups can be combined into topology groups top, up, lateral, down and bottom for the event classification of IV events.}
\label{PMTgrouping}
\end{figure}

\noindent The trigger and timing system for the Double Chooz experiment consists of three TBs and one TMB as shown in Figure \ref{TriggerSchematic}. Two TBs are used for the signals from the ID. Each ID TB is connected to half of the 390 PMTs using 13 of the 18 input channels\footnote{This leads to 12 groups of 16 PMTs and one with only three}. The PMT inputs are grouped in an alternating way such that each TB observes the same detector volume (see Figure \ref{IDgrouping}). This was motivated by two reasons \cite{DissAC}. Using two identical boards allows to determine the trigger efficiency by cross comparison. Furthermore, redundancy and high robustness is introduced to the system. If one board fails no events are lost and the failure can be detected offline. The third TB is used for the signals from the IV. Each of the 18 group signals consists of PMTs of a certain region of the IV vessel (see Figure \ref{IVgrouping}).

\noindent All the VME cards are hosted in a VME64x crate and are controlled by a VME based computer system MVME3100 (Emerson Network Power). The same computer is used in all other VME crates of the experiment.\\

\subsection{Trigger Logic for the Double Chooz experiment}
\label{DClogic}

The trigger logic is a combination of the sum thresholds and a multiplicity condition on the single group thresholds. Whenever the read out condition of one of the three TBs is fulfilled or an external trigger signal is active, a trigger will be generated by the TMB. In this section we will describe the setup of the trigger system and the resulting Trigger Word.\\

\subsubsection{ID Trigger Boards}
\label{ID_settings}
For the ID TBs four sum thresholds are called the ``prescaled'', the ``neutrino like'', the ``neutron like'' and the ``muon like'' threshold. The prescaled threshold is lower than the neutrino threshold such that it delivers a rate of 1000/s scaled by a factor of 1/1000 (on the TMB) leading to a total rate of 1/s of prescale triggers. The ``neutrino like'' threshold is the physics threshold of the ID. It is set to approximately 350 keV well below the minimum energy of the inverse beta decay of 1.02 MeV by which neutrinos are detected \cite{DCFirstPub}. It runs at a rate of approximately 100/s. The two highest thresholds are used as flags for the event classification and set to approximately 6 MeV for ``neutron like'' events and 50 MeV for muons, respectively. The group thresholds are set to 80$\,\%$ of the ``neutrino like'' threshold. For a trigger at least 2 out of the 13 groups have to fire. For the prescaled trigger no multiplicity condition is applied.

\subsubsection{IV Trigger Board}
\label{IV_settings}
For the IV TB three sum thresholds are called the ``prescaled'', the ``neutron like'' and the ``muon like'' threshold. The prescaled threshold is set similarly to the ID. It creates 1000/s triggers which are scaled by a factor of 1/1000 leading to a total rate of 1/s prescale triggers. Because of the geometrical properties of the IV: the inhomogeneous distributed PMTs and the anisotropic light propagation the IV TB is treated differently than the ID TB boards.
From a Threshold-scan we determined a rate spectrum for each discriminator of the IV TB (cf. section \ref{threshscan}). With those spectra it is possible to distinguish between regions (threshold values) where the rate is dominated by radioactivity (cf. section \ref{IVscans}). The ``neutron like'' threshold is set right above the value where the rates are dominated by radioactive background (``neutron spot'', see Figure \ref{ThresholdScanDC_IV}). For the group discriminators the thresholds have been calculated taking into account the number of PMTs belonging to the corresponding group (cf. Figure \ref{IVgrouping}). The ``neutron like'' threshold is the physics threshold of the IV. It is set to approximately 10 MeV. There is no multiplicity condition for these triggers. The ``muon like'' threshold is set to approximately 50 MeV and combined with a multiplicity condition of at least 10 active groups.

\subsubsection{Muon Event Categorization}

\begin{figure}[htb]
\centering
\includegraphics[angle=0,width=0.45\textwidth]{{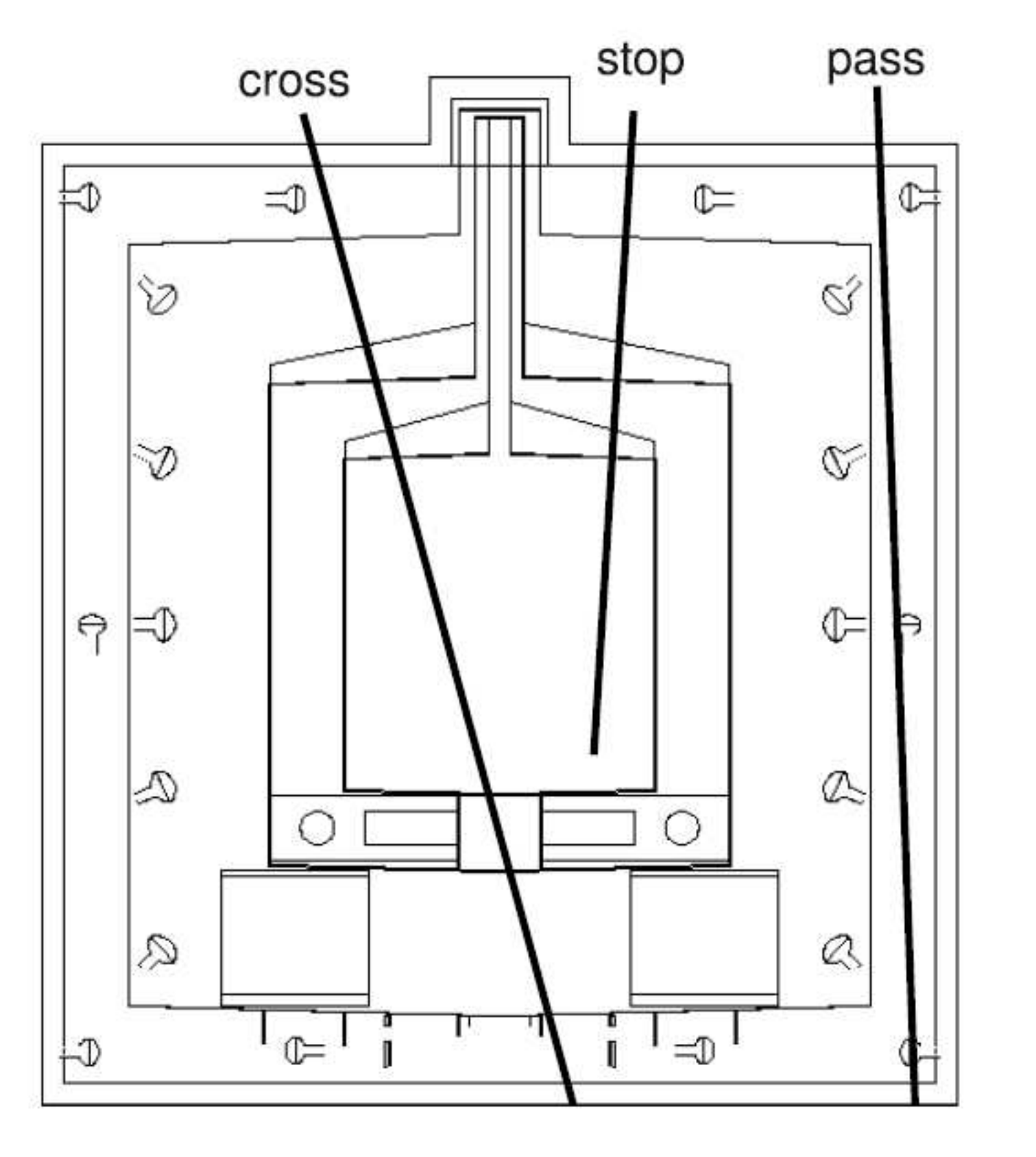}}
\caption{The muon categories considered by the Inner Veto Muon Pattern Recognition}
\label{IVMPRprinc}
\end{figure}

\noindent Simulations have shown that it is possible to achieve a rough muon event categorization IVMPR (\textbf{I}nner \textbf{V}eto \textbf{M}uon \textbf{P}attern \textbf{R}ecognition) at the trigger level. Muon events can be distinguished in three different classes (see Figure \ref{IVMPRprinc}): muons which stop inside the ID (stopping), muons which cross the IV without entering the ID (passing) and muons which cross the ID and the IV (crossing).\\
First the ``topology'' bit is derived by a logical OR of three multiplicity conditions in the IV: at least three active groups of the lateral group OR at least one active group in the down area OR an active bottom group (see Figure \ref{IVgrouping}). I.e. for $\overline{\rm topology}$ only groups belonging to the top and the up are allowed to be active. The IVMPR logic is the following:

\begin{itemize}
\item \textbf{passing muon: } $ \rm ``muon \, like \,(IV)'' \,  \land \, \overline{muon \, like\, (ID)}$
\item \textbf{stopping muon: } $ \rm ``muon \, like \,(IV)'' \,  \land \, muon \, like\, (ID)\, \land \,\overline{topology}$
\item \textbf{crossing muon: } $ \rm ``muon \, like \,(IV)''  \, \land \, muon \, like\, (ID)\, \land \,topology$
\end{itemize}

\noindent The IVMPR categorization is recorded in the Trigger Word.

\subsubsection{The Double Chooz Trigger Word}
\label{DCTW}
On the TMB all signals from the TBs and external trigger sources, e.g. from the calibration systems \cite{DCFirstPub} are processed via separate CAMs into the TW (see Figure \ref{TMB_scheme}). The scaling of the prescaled triggers is also done on the TMB.\\
\noindent Except for the fixed rate trigger, which is set to a rate of 1/s, all special triggers of the high level logic are disabled. After each trigger a forced dead time of 128 ns is set to prevent the event window of the FADCs of two consecutive events from overlapping (for the data published so far an event window is 256 ns long \cite{DC2ndPub}).\\

\noindent The configuration of the TW is the following:

\begin{itemize}
\item \textbf{Bit 0-3: }TB A: prescale, neutrino like (read out), neutron like (flag), muon like (flag).
\item \textbf{Bit 6-9: }TB B: prescale, neutrino like (read out), neutron like (flag), muon like (flag).
\item \textbf{Bit 12-17: }TB IV: prescale, neutron like (read out), muon like (flag), passing muon (flag), stopping muon (flag), crossing muon (flag).
\item \textbf{Bit 20-25: }External triggers from various calibration systems \cite{DC2ndPub}, trigger dead time monitor system (cf. section \ref{DTM}) and the OV (flag).
\item \textbf{Bit 28: } Fixed Rate trigger
\end{itemize}

\noindent Important for the near detector is the ability to reduce the amount of stored data based on the event classification of the TW (cf. section \ref{OutputsOfTS}). For the data published up to now, with only the far detector, this option has not been used. However, it is planned to test this method with the far detector in the near future.

\subsection{Output of the Trigger and Timing System}
\label{OutputsOfTS}

In order to distribute the output signals to the FADC crates several custom Fan-Outs\footnote{embedded on VME cards, such they can be plugged into the DAQ crates} are part of the trigger and timing system (see Figure \ref{TriggerSchematic}). The output signals of the trigger and timing system (see Figure \ref{TMB_scheme} and \ref{TriggerSchematic}) are explained below:

\begin{itemize}
\item \textbf{Trigger 1 signal (TR1):} It triggers the readout of the detector. It is a standard NIM pulse. It is distributed to all FADC modules via several dedicated fan-out modules.
\item \textbf{Clock signal (Clk):} The trigger system provides a common 62.5 MHz (16ns period) clock to all components for synchronous data taking. The clock signal is based on LVDS (\textbf{L}ow \textbf{V}oltage \textbf{D}ifferential \textbf{S}ignaling). It is distributed to all components via several dedicated Fan-Out modules.
\item \textbf{OV sync signal:} This signal can be used to synchronize systems which are using a separate clock. It is used to synchronize the OV. The OV sync signal is sent every 68.72 s (0.015 Hz). It is derived from the Clk signal.
\item \textbf{Inhibit signal (INH):} To guarantee a synchronous start of all components the trigger system provides an inhibit signal (standard NIM pulse). While this signal is active no data will be collected. Upon deactivation, all components start running.
\item \textbf{Trigger Word (TW):} Is a classification of each triggered event. This can be used to reduce the amount of data stored. The event classification is coded in the 32 bit TW and distributed from a 34-pin connector socket (LVDS) via dedicated fan-outs to the FADC crates, in time with the trigger signal (cf. section \ref{TMB}). The FADCs can read only 16 LVDS signals so that the 32 bit TW had to be split into two FADCs per crate.
\item \textbf{Event Number (EvNo):} The trigger system provides a counter which is incremented with every trigger. This 32 bit counter (EvNo) is distributed via LVDS Fan-Outs to each FADC crate synchronous with the trigger. The EvNo sent from the trigger system is compared to the internal event counter in the FADC to validate the data before it is written to disk.
\end{itemize}

\section{Commissioning and Tests of the Double Chooz Trigger System}

The installation of the electronic components was done in April 2010. The construction of the far detector was completed at the end of 2010. During this period tests of the subsystems have been performed \cite{CK_Thesis}. The trigger system was used to monitor the detector behavior e.g. during the filling of the detector. A detailed summary of the development of the trigger system and its earlier tests can be found in \cite{BerndPhDT}. 

\subsection{Discriminators Performance}

\begin{figure}[htb]
\centering
\subfloat[]{
\includegraphics[angle=0,width=0.45\textwidth]{{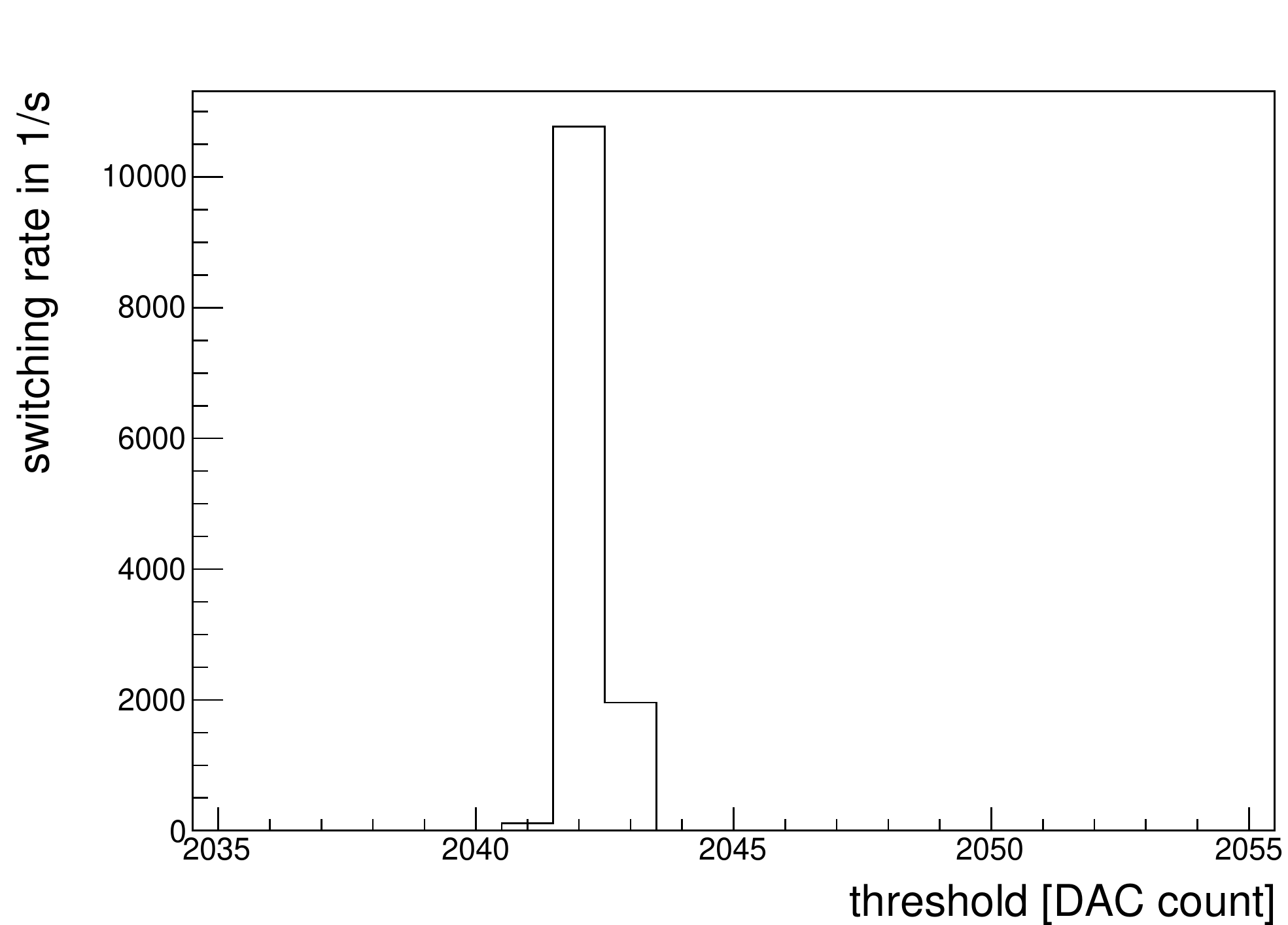}}
}
\subfloat[]{
\includegraphics[angle=0,width=0.45\textwidth]{{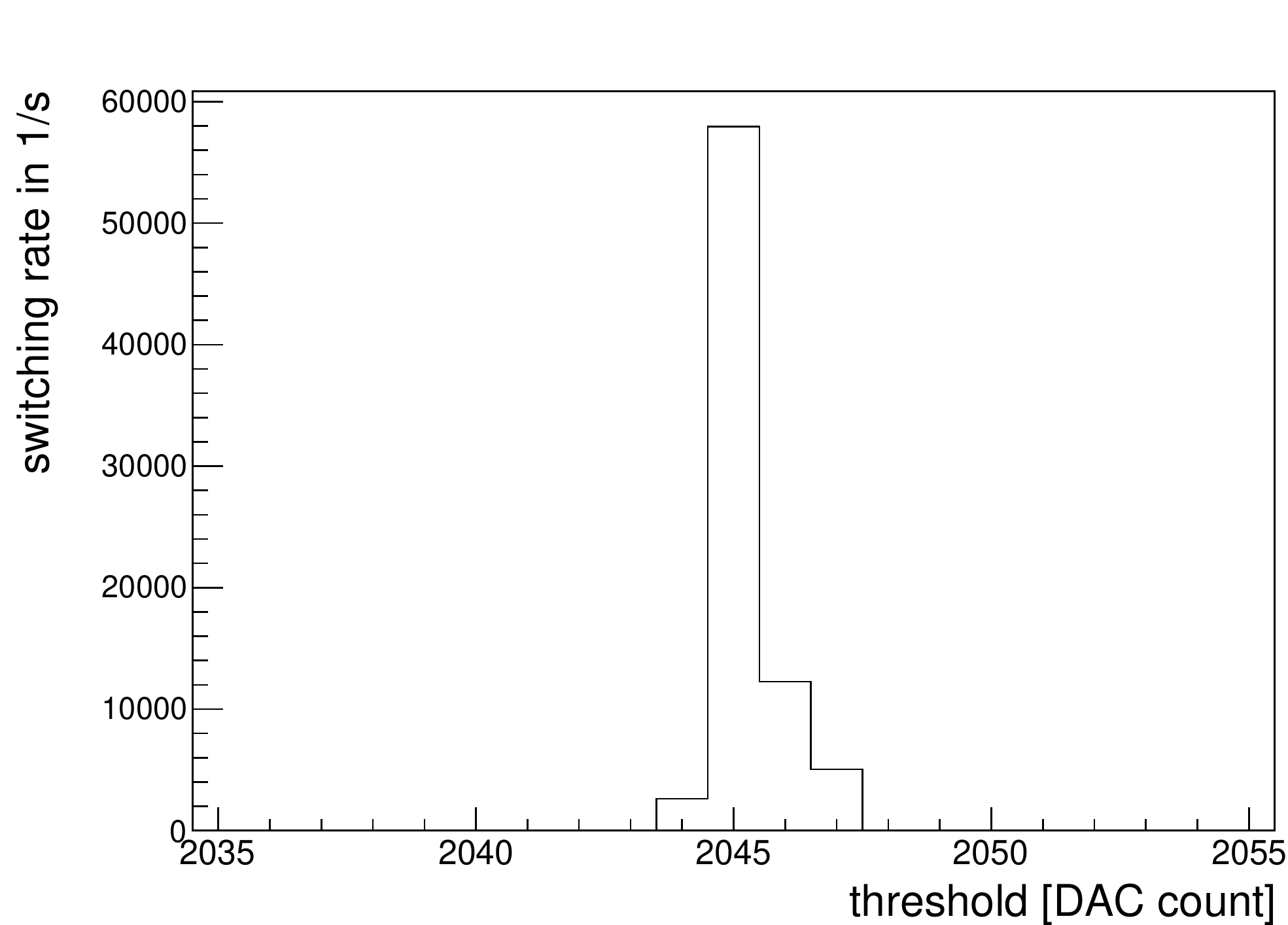}}
}
\caption{Example of a Threshold-scan without an input connected to the TB. (a): Sum discriminator (b): Group discriminator}
\label{BaselineScan}
\end{figure}

The performance of the discriminators is critical for the overall performance of the trigger system. Before the system was installed at Chooz the discriminators have been calibrated carefully. Crosstalk, internal electronic noise and linearity of the discriminators was studied.\\
To estimate the internal electronic noise of the TB we used ``Threshold-scans'' (cf. section \ref{threshscan}). The region around each discriminator baseline was scanned count by count. For each step the switching rate of the discriminator (IRC value, cf. section \ref{TBdata}) was measured. Figure \ref{BaselineScan} shows the result for a sum and a single channel discriminator.

\noindent The different total of the baseline are related to the different dynamic ranges and therefore different steps of the thresholds of the group and the sum discriminators. The values for one TB are listed in Table \ref{Resolutions_T}. The discriminators start switching if the threshold is near the baseline and noise creates triggers. The distributions in Figure \ref{BaselineScan} give an estimation of the internal electronic noise of the TBs. Both examples show that the level of electronic noise is negligible (less than one DAC).\\

\begin{table}
\centering
\begin{tabular}[ht]{|c|c|}
  \hline
  channel &  resolution [mV/DAC] \\
  \hline\hline
   single input (A/B)    &  $\approx$ 0.62 \\
   SUM A (neutron like)  &  9.35 \\
   SUM B (muon like)     & 18.18 \\
   SUM C (prescale)      &  5.85 \\
   SUM D (neutrino like) &  9.35 \\
\hline
\end{tabular}
\caption{Threshold steps of a Trigger Board in mV input signal per DAC count. The single channel value is the average value for all 18 group inputs.\label{Resolutions_T}}
\end{table}

\begin{figure}[htb]
\centering
\subfloat[]{
\includegraphics[angle=0,width=0.45\textwidth]{{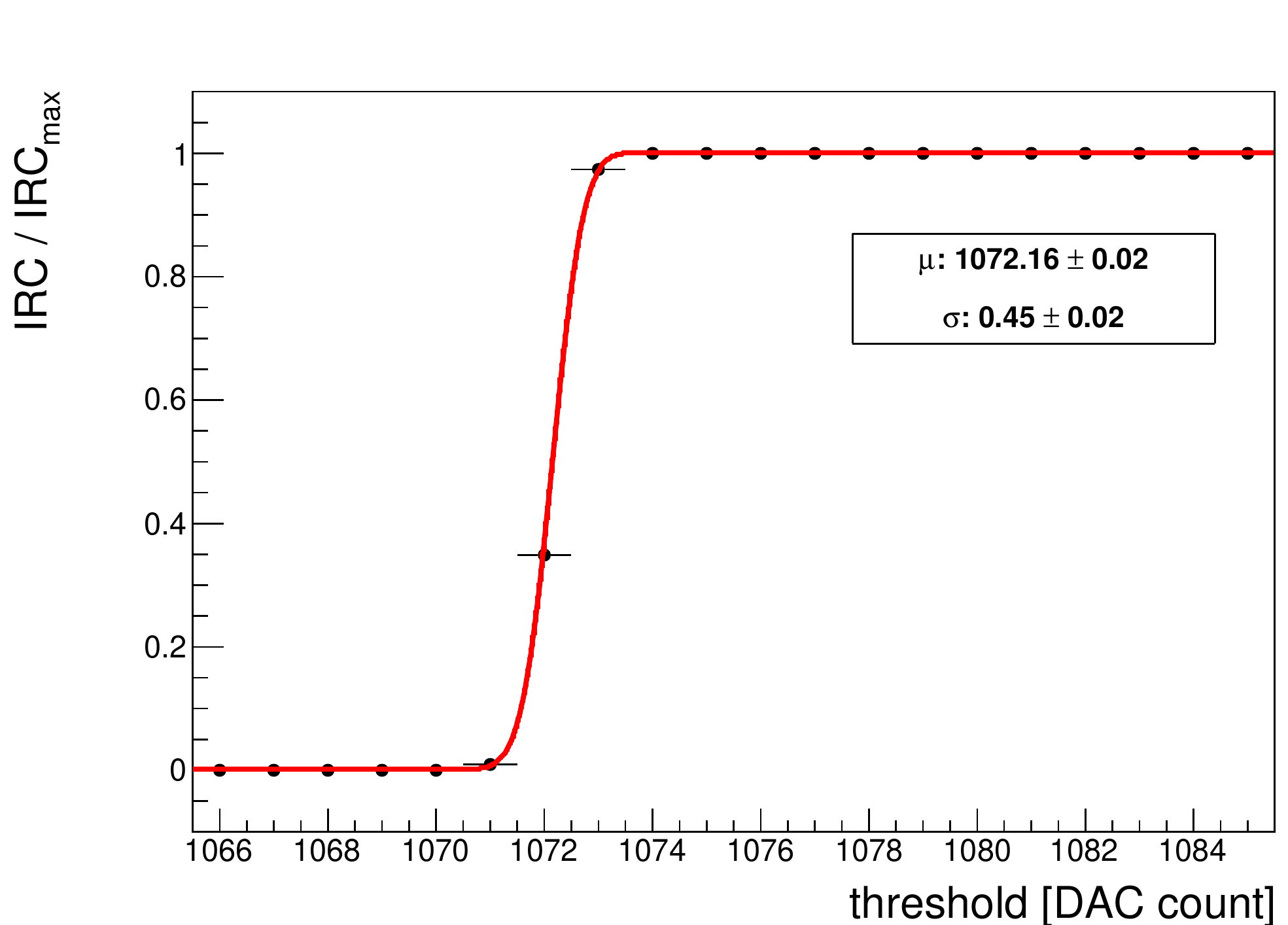}}
\label{ThresholdLin}
}
\subfloat[]{
\includegraphics[angle=0,width=0.45\textwidth]{{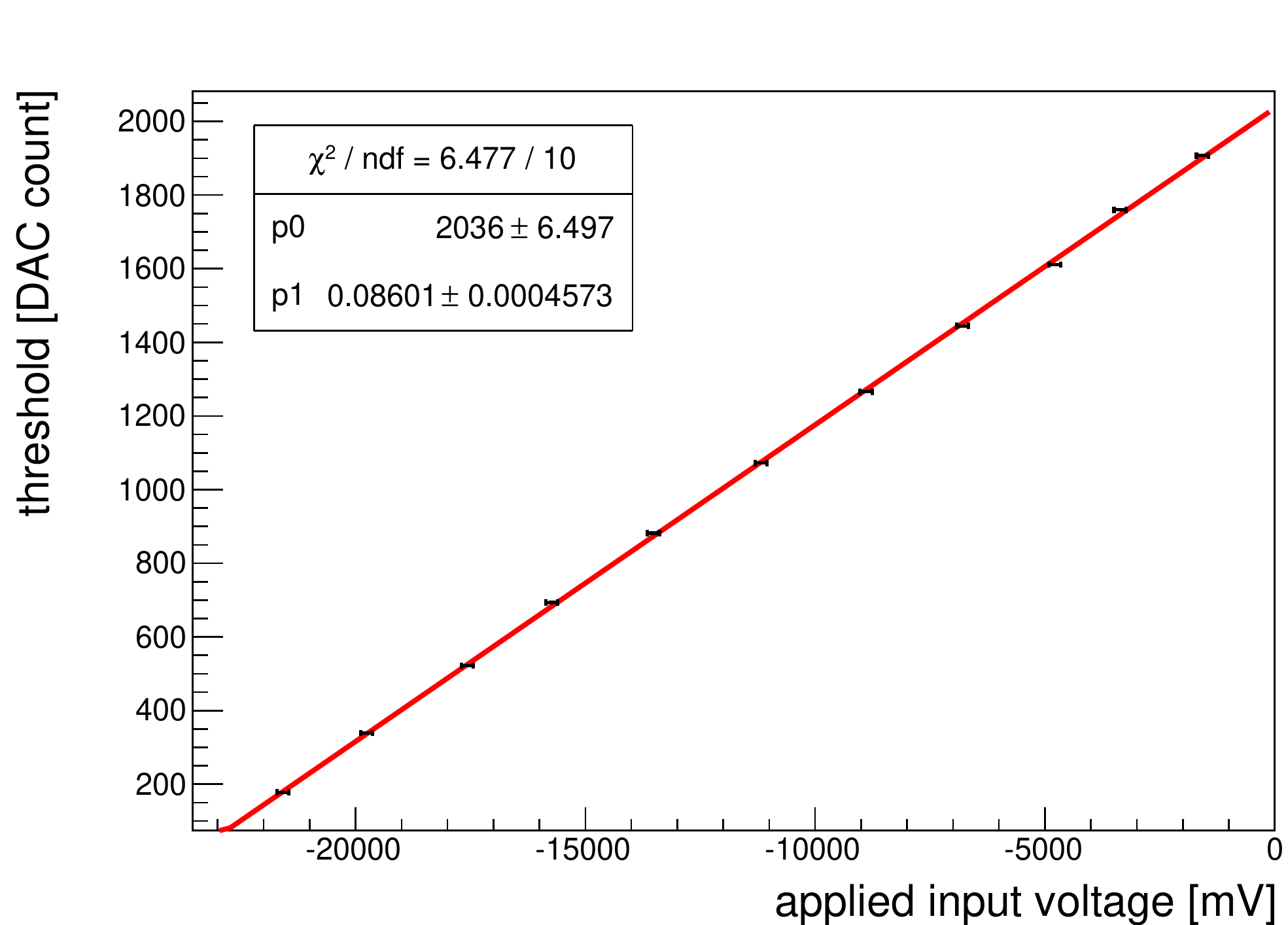}}
\label{LinFit}
}
\caption{(a): Threshold-scan distribution for a negative input pulse and fitted error function. (b): Linear fit of the threshold values determined for several input pulse heights. NOTE: A DAC value of 0 corresponds to the maximal negative threshold, a value of around 2000 to zero input and values of up to 4096 would discriminate positive inputs} 
\label{BaselineScanLin}
\end{figure}

\noindent A similar measurement can be used to estimate the crosstalk between different channels. Here a signal was connected to one group only and the IRC (cf. section \ref{TBdata}) values of adjacent discriminators have been measured under the influence of the the discriminator switching of nearby \cite{BerndPhDT}. This measurements showed that there is only negligible crosstalk between channels (less than one DAC).\\
\noindent The linearity of the discriminators was verified with a pulse generator using rectangular pulses with different pulse heights. For each pulse height the corresponding threshold was determined by a Threshold-scan (see Figure \ref{ThresholdLin}).

\noindent The distribution can be described by an error function which is fitted to determine the threshold $\mu$ for a certain pulse height.

\begin{equation}
f(DAC) = \frac{\alpha}{2}\left(1 + erf \left( \frac{(DAC - \mu)}{\sqrt{2 }\sigma}\right) \right)
\label{ErrorFkt}
\end{equation}

\begin{center}
Where $\alpha$ stands for the total number of pulses. $\mu$ and $\sigma$ are the threshold and threshold resolution, respectively.
\end{center}

\noindent All discriminators show a similar linear behavior over their dynamic range as the one shown in Figure \ref{LinFit}. A temperature dependence of the discriminators has not been observed because the electronics of Double Chooz is constantly cooled by a ventilation system in the underground lab.

\subsection{Clock Stability}

The total run time of the experimental data sets is calculated from the clock counter of the TMB (cf. section \ref{TMBdata}). The clock counter value is continuously compared with the time difference counter (cf. section \ref{TBdata}) of the TBs and the clock counters of the $\nu$-FADCs to ensure synchronicity of the cards. The reliability of the TMB clock has been verified with a custom GPS board. This board is a modified TMB and is planed to be installed and used for the Double Chooz experiment in the future. It receives a GPS time stamp every 200 ms (5 Hz) from a Lea4T GPS module. The time difference between receiving the GPS signal and the trigger signal from the trigger system, is measured with a counter in steps of 8 ns. Whenever a trigger signal is received the GPS time stamp and the counter value is stored in the FIFO of the GPS board, thus the actual time stamp for the trigger signal can be calculated afterwards.\\
In several laboratory tests with 24 hours of run time each, the run length calculated from the TMB (using the clock counter, cf. Section \ref{TMBdata}) has been compared with the run length stated by the GPS clock. In order to create triggers the fixed rate trigger of the TMB has been used. These measurements lead to an uncertainty of 0.018 s/h in the run length calculated from the TMB clock, which is well below the manufacture's specification of 0.072 s/h \cite{OscillatorRef}. The manufacture's specification also covers aging of the oscillator.\\
\noindent The time stamps needed for analysis e.g. search for delayed coincidences of physics events like the inverse beta decay or the comparison to the reactor power history are in the order of $\mathcal {O}(\mu s)$ and $\mathcal{O} (\min)$, respectively. Therefore the precision and stability of the clock is well below the requirements of the experiment.\\
The planned installation of the custom GPS board in the experiment provides a potential application to contribute to a global campaign, such as SNEWS \cite{SNEWS} for the detection of neutrinos from supernova explosions.

\subsection{Threshold-scans for Commissioning the Far Detector Trigger}
\label{threshscan}

The trigger system can be operated independently from the other DAQ components. The trigger system allows an almost instantaneous monitoring of the detector performance because of the simple and robust structure of its data. This has been useful during the filling of the detector and the commissioning of the DAQ.\\
Threshold-scans were taken regularly. The distribution of the IRC values shows a cumulative rate spectrum of the corresponding discriminator and detector. Note, that the trigger system is triggering on negative pulses, therefore the pulse heights increase from the right to the left in all plots shown in this section, with the dotted line indicating zero input around 2040.

\subsubsection{Inner Detector}
\label{IDscans}

\begin{figure}[htb]
\centering
\includegraphics[angle=0,width=0.9\textwidth]{{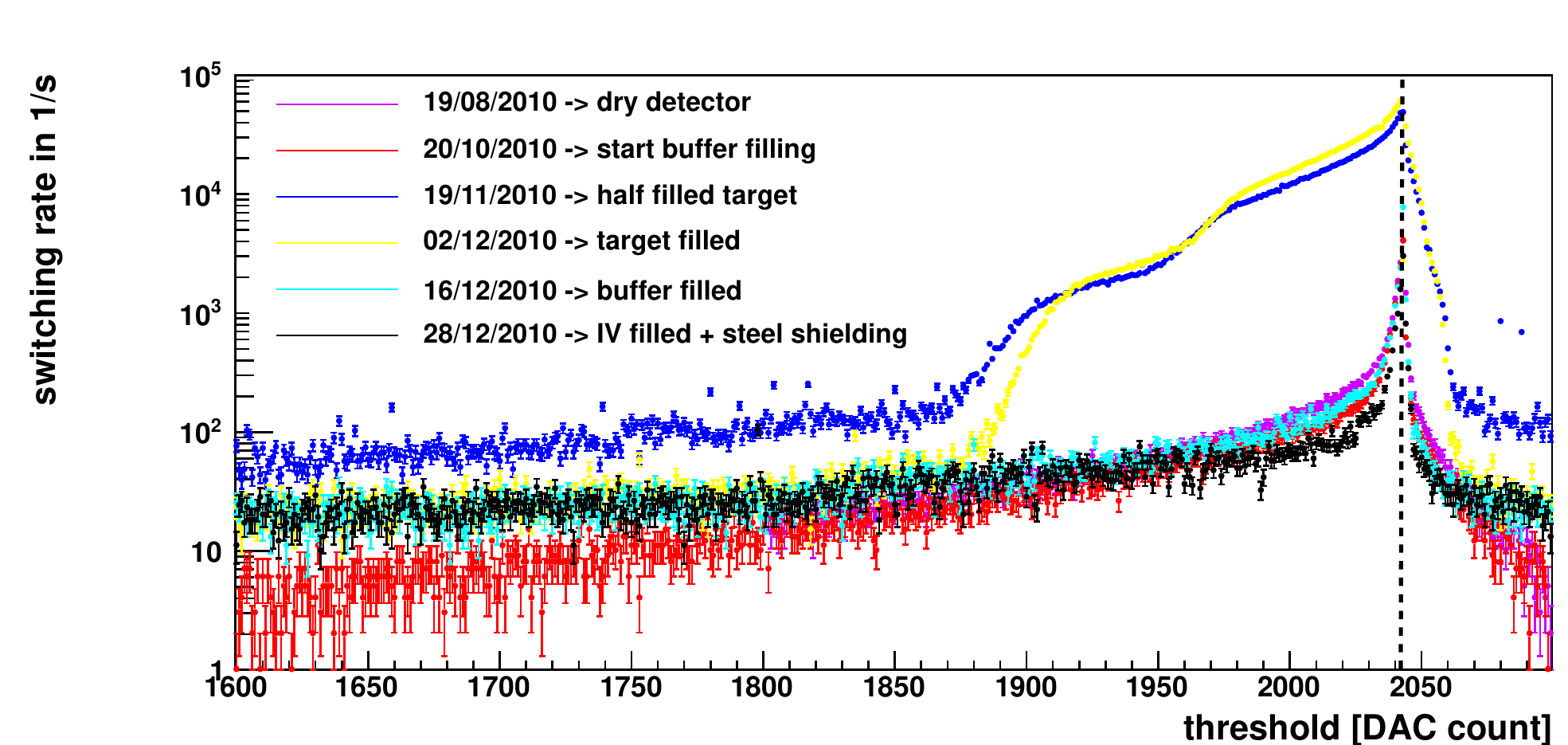}}
\caption{Threshold-scan distributions of the ``neutrino like'' threshold of an ID TB at the far detector. The plotted distributions cover the whole filling period of the far detector (12.10.10-13.12.10) and show how the trigger behavior changed during that time. Further, a distribution taken right after the electronics installation (with a dry detector) is shown. The dotted line indicates the baseline position for the shown threshold.}
\label{ThresholdScanDC_ID}
\end{figure}

\begin{figure}[tb]
\centering
\includegraphics[angle=0,width=0.9\textwidth]{{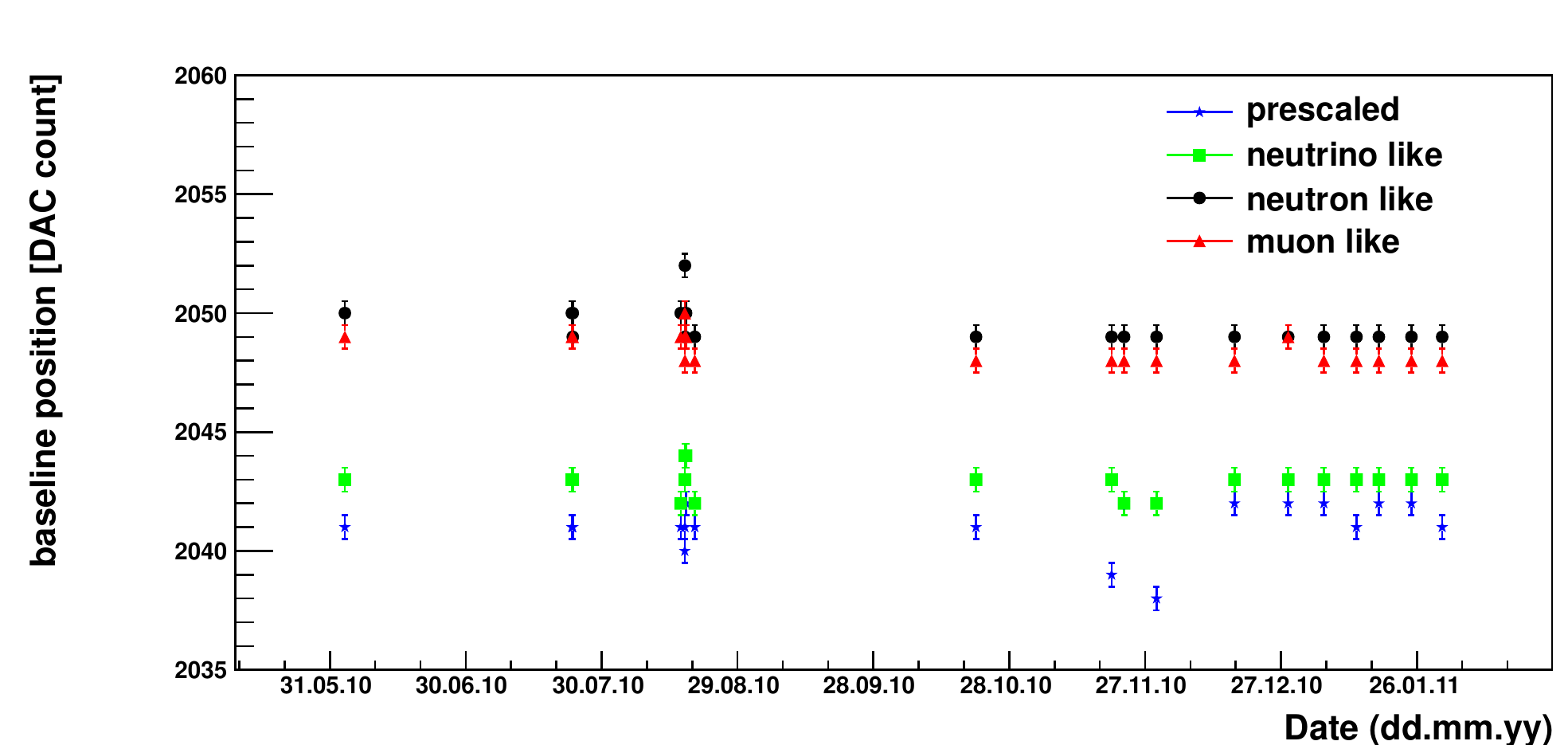}}
\caption{Baseline positions shown for the sum discriminators of one ID TB for the time period between installation and start of data taking. The variations are below one DAC count over one year.}
\label{BaselineStab}
\end{figure}

\noindent The distributions in Figure \ref{ThresholdScanDC_ID} show how the IRC changes during the filling operation. The violet data points show the discriminator response when the detector was empty. The red data points have been measured when the liquid levels reached the buffer volume \footnote{Note, that due to the delicate acrylics of the inner volume vessels all volumes were filled simultaneously, allowing only a few centimeter deviation between the different liquid levels. If for example the target was completely filled the liquid levels of the buffer and IV volumes are at the upper boundary of the target as well.}. The slightly higher rates measured with a dry detector can be explained by a higher dark count rate after turning on the high voltage \cite{PMTpaper} and radioactivity from the rock below the detector. With the increase of the liquid levels inside the detector an increase of the rates is observable. When the liquid level reaches the target vessel (filled with scintillator) a structure becomes visible around DAC 1890 and DAC 1960, respectively. This structure can be explained by the natural radioactivity from the detector and the surroundings around the detector and the PMT cathodes. The highest rates have been observed when the target was half filled (blue data points). With further increasing of the liquid levels a decrease of the rates has been observed. When the buffer volume was filled completely the structure in the rate spectrum disappeared. This observation and the reduction of the rates at higher filling levels show that the buffer volume act as a passive shielding for the innermost volumes. The black data points show the switching rate after the whole detector was filled and the steel shielding was installed. The rates for small amplitudes have decreased further because of the passive shielding of the IV liquid and the steel shielding. The asymmetry of the distributions around the baseline can be explained by the response of the ISS circuit to positive signals (cf. Figure \ref{ISS}).\\
After the completion of the filling one can observe that the rate at high amplitudes (which are corresponding to high energies cf. section \ref{DAQandDetector}) converge to the expected muon rate of approximately 10/s \cite{DCProposal}.
After the installation the Threshold-scan distribution has been regularly monitored and found to be stable. The variations between different measurements are consistent within the statistical fluctuations. The stability of the electronics, in particular the stability of the input baseline, can be monitored from the maximum in the distributions. Figure \ref{BaselineStab} shows that the baseline for all discriminators has been stable (variations below one DAC) since the electronics installation, demonstrating the stable behavior of detector and electronics.

\subsubsection{Inner Veto}
\label{IVscans}

\begin{figure}[htb]
\centering
\includegraphics[angle=0,width=0.9\textwidth]{{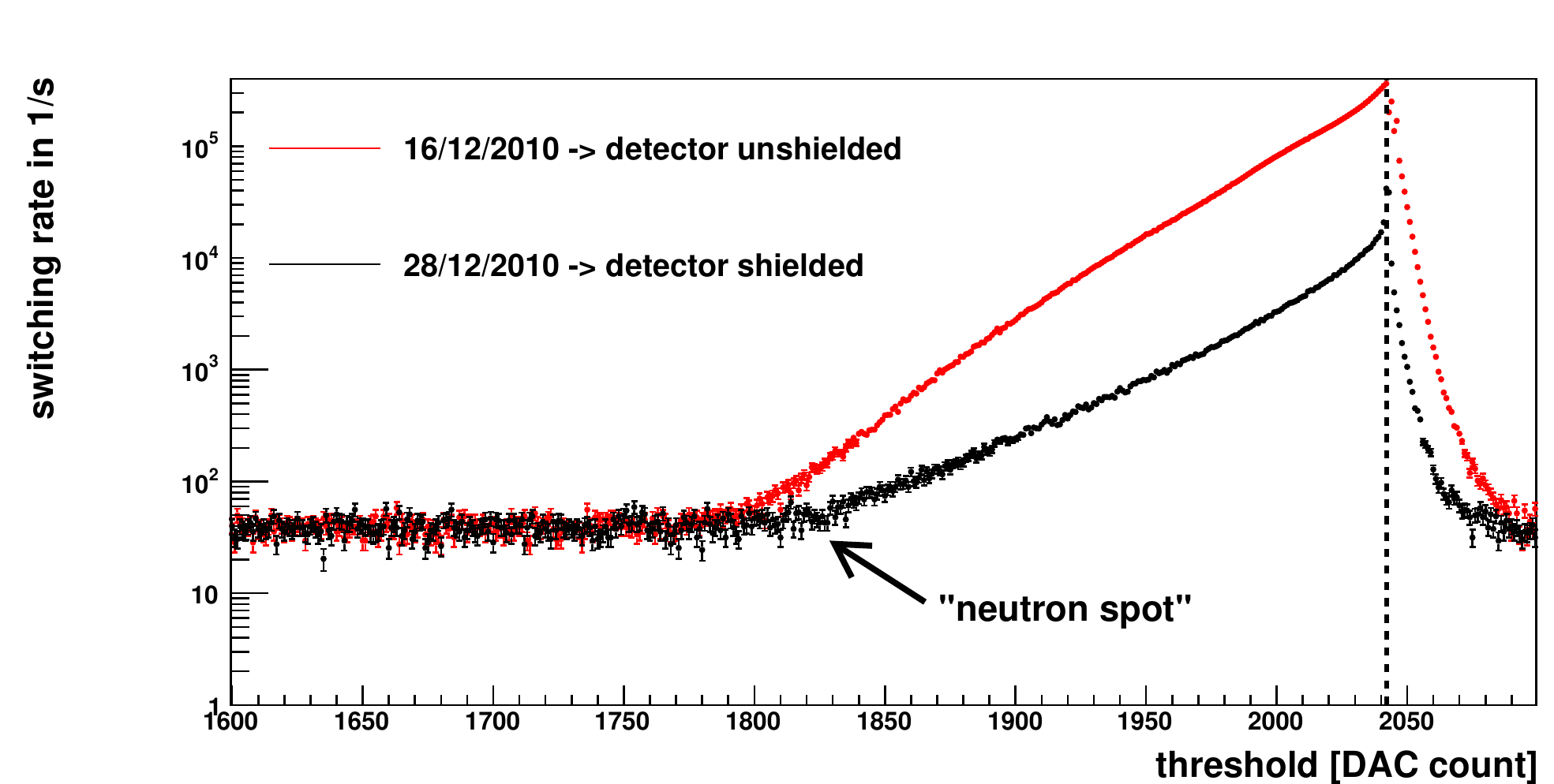}}
\caption{Threshold-scan distributions of the ``neutron like'' threshold of the IV TB at the far detector. The red and black data points show the discriminator response before and after the installation of the upper steel shielding. The dotted line indicates the baseline position for the shown threshold.}
\label{ThresholdScanDC_IV}
\end{figure}

\begin{figure}[htb]
\centering
\subfloat[]{
\includegraphics[angle=0,width=0.45\textwidth]{{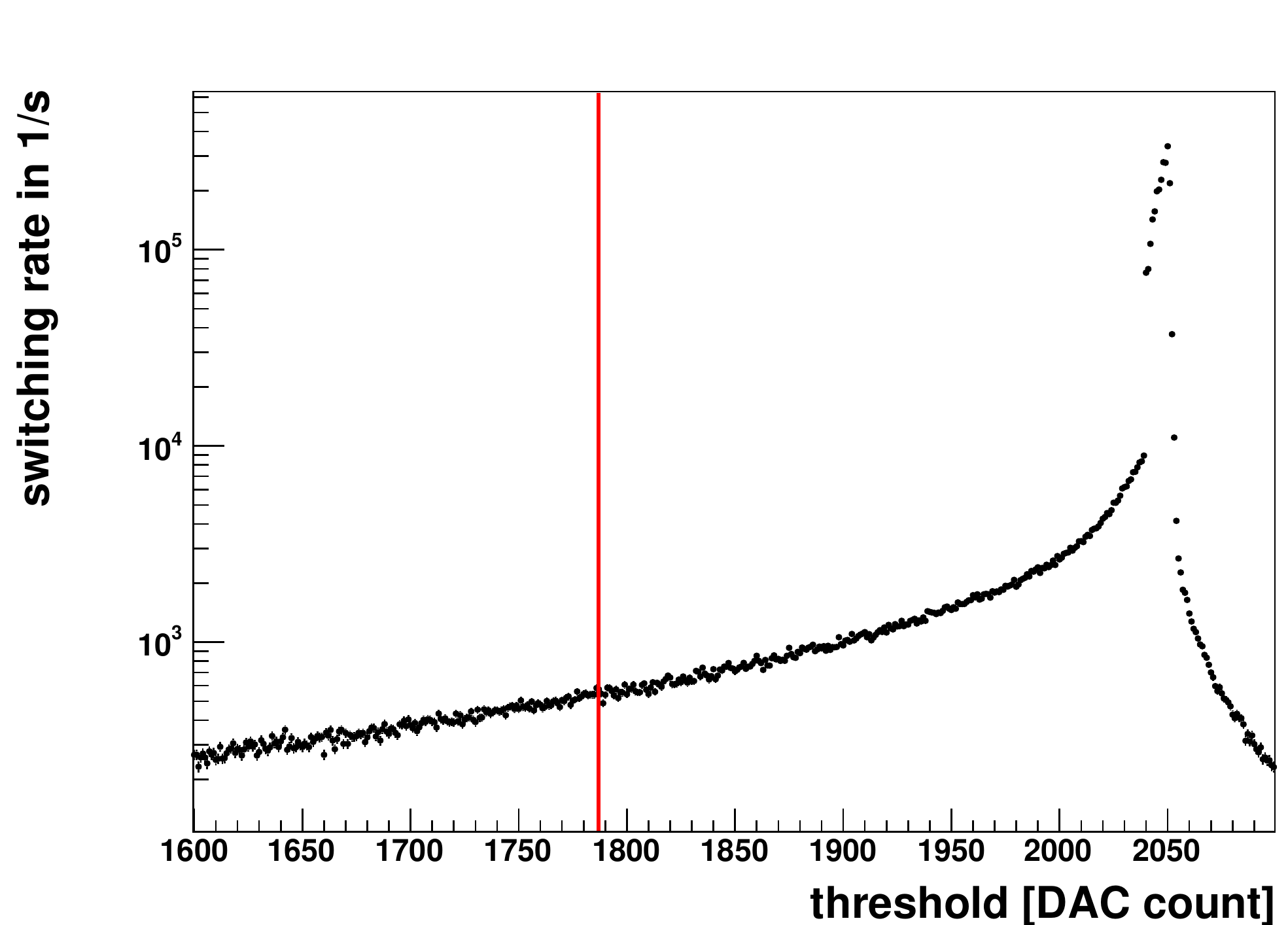}}
\label{TopGroup}
}
\subfloat[]{
\includegraphics[angle=0,width=0.45\textwidth]{{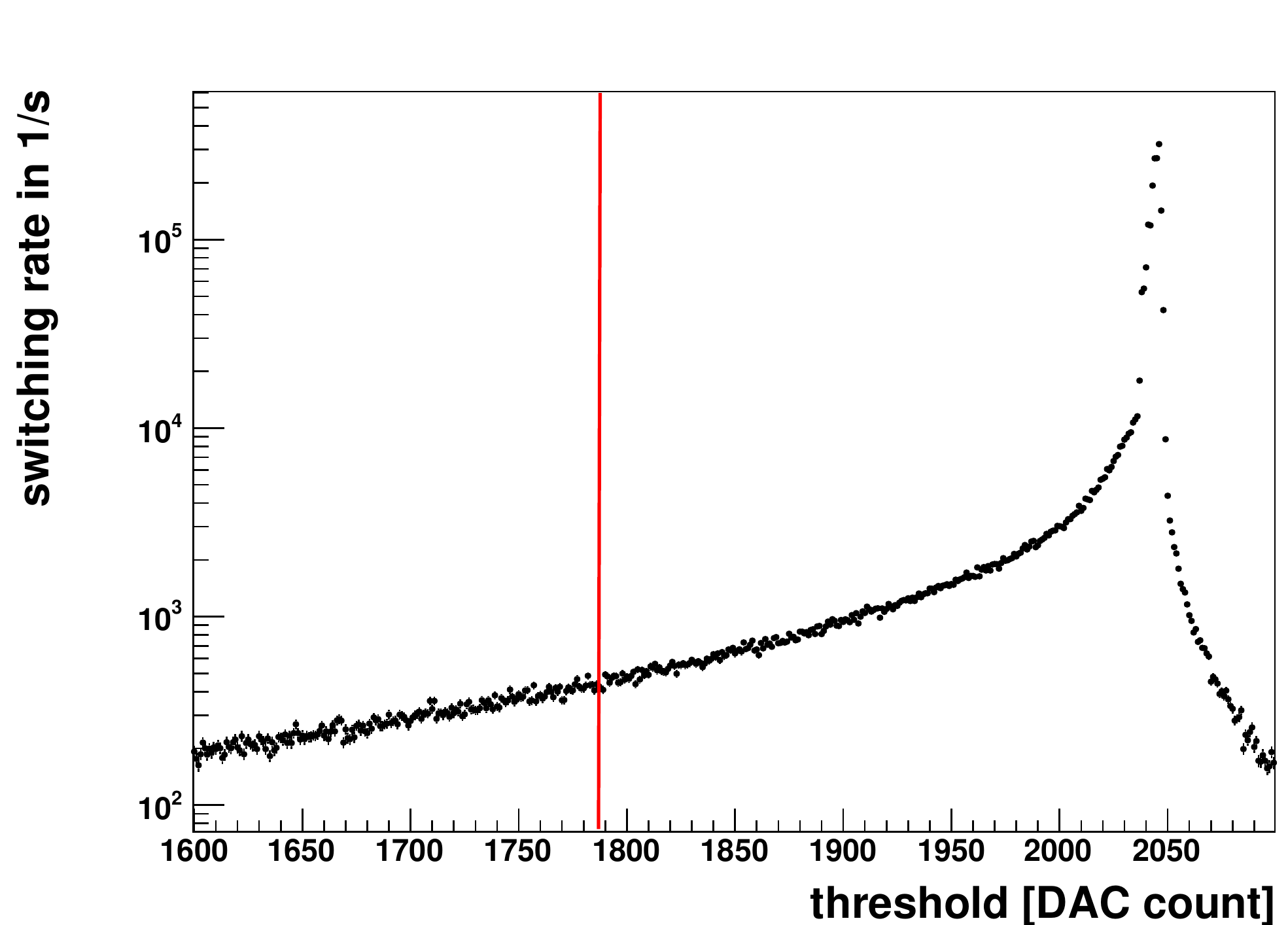}}
\label{BottomGroup}
}
\caption{Examples of Threshold-scan distributions for the IV TB group discriminators. Red lines indicate the thresholds position. (a): Top group. (b): Bottom group.}
\label{Thresholdscans_IV}
\end{figure}

\noindent The distributions in Figure \ref{ThresholdScanDC_IV} show the IRC of the ``neutron like'' threshold of the IV TB at the far detector before (red) and after (black) the steel shielding was completed. The steel shielding above the detector was installed after the detector was filled. The decreased rates for smaller amplitudes after the shield installation demonstrate how efficiently the steel shielding surrounding the detector protects it from radioactivity from the outside. At high amplitudes (energies cf. section \ref{DAQandDetector}) both distributions converge to the expected muon rate (approximately 30/s \cite{DCProposal}) for the IV at the far detector site.\\
For the IV, the Threshold-scans have also been used to define the threshold settings (cf. section \ref{IV_settings}). Both data sets in Figure \ref{ThresholdScanDC_IV} show a spot were the rates converge to a constant rate. Between this spot (``neutron spot'') and the baseline position the rates are dominated by radioactivity from outside the detector. The ``neutron like'' threshold for the IV has been set to the so called ``neutron spot'' (see Figure \ref{ThresholdScanDC_IV}). For the group discriminators the threshold has been calculated by scaling the sum signal amplitude at the ``neutron spot'' according to the number of PMTs of a particular group (see Figure \ref{IVgrouping}). Figure \ref{Thresholdscans_IV} shows examples of Threshold-scan distributions for the IV groups. The shape of the distributions are rather similar for all group discriminators. Small differences in the rates at the calculated ``neutron spot'' threshold (red line) can be explained by the number of PMTs connected to the particular group and the topological position of the groups (cf. Figure \ref{IVgrouping}).

\subsection{Multiplicity}

\begin{figure}[tb]
\centering
\includegraphics[angle=0,width=0.9\textwidth]{{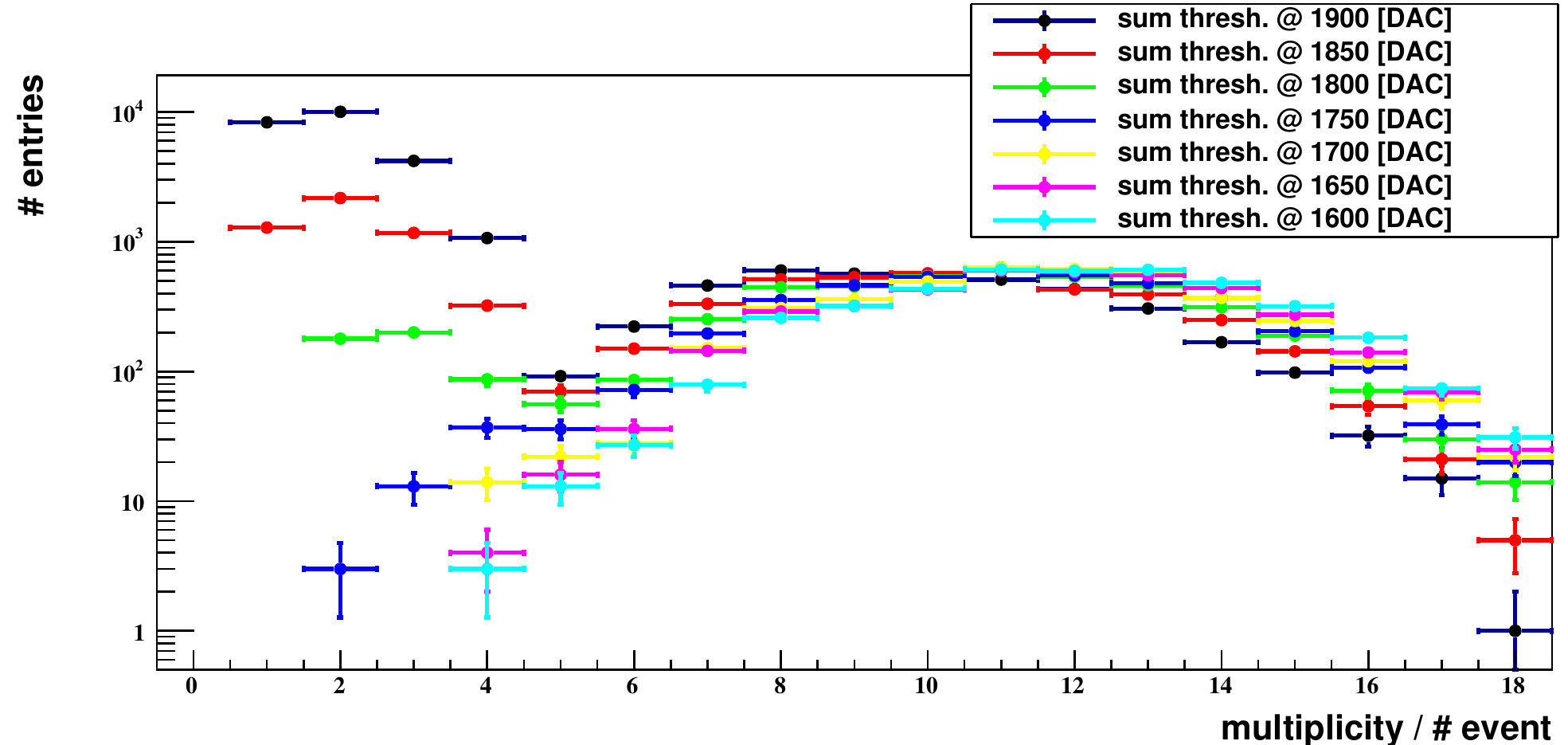}}
\caption{Multiplicity study for the IV TB. The group thresholds have been fixed to the ``neutron spot'' equivalent of the sum threshold of the IV. The distributions show the number of active group channels (multiplicity) per event.}
\label{Multiplicity_vs_SUM}
\end{figure}

During the commissioning of the Trigger System studies on the multiplicity (number of active groups) have been done. Here we discuss the results for the IV TB as an example. Figure \ref{Multiplicity_vs_SUM} shows how the number of active groups changes with an increase of the sum threshold\footnote{Corresponding to an increasingly negative voltage and hence a smaller DAC count}. Note, that all sum threshold values for this studies are below the thresholds for pure muons.

\begin{figure}[htb]
\centering
\includegraphics[angle=0,width=0.9\textwidth]{{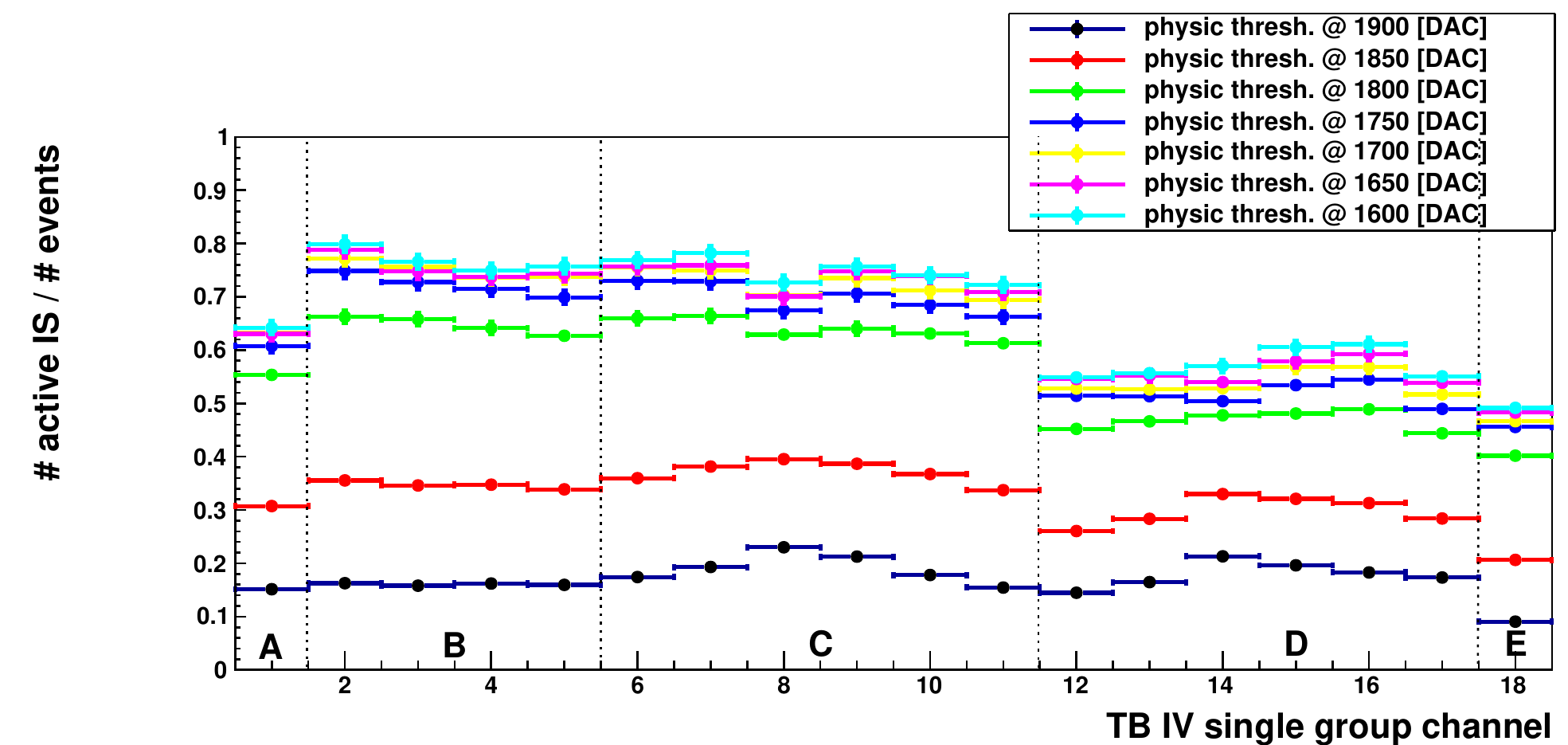}}
\caption[]{Multiplicity study for the IV TB. The number of times were the IS was active divided by the total number of events recorded within 100 s for different sum threshold levels is plotted for each TB IV group channel. Dotted lines indicate different topology regions of the IV (cf. Figure \ref{IVgrouping}). A: Top, B: Up, C: Lateral, D: Down, E: Bottom.}
\label{ISvsSUM_IV}
\end{figure}

\noindent One can see that for a low sum threshold (e.g. 1900 [DAC]) the rate of low multiplicity is high because the IV mainly triggers on the localized energy deposition from radioactivity from the outside. With the increase of the sum threshold the mean multiplicity is also increased. This correlation between the sum threshold value and the resulting mean multiplicity shows that a certain sum threshold could be replaced by a corresponding multiplicity condition, i.e. both conditions are redundant. A low sum threshold combined by a logical AND with a high multiplicity condition would effectively increase the physic threshold to a higher level then the actual set sum threshold level. Even if the sum threshold is set to a low value the detector would be triggering only on high energies because of the multiplicity condition.\\
The physics threshold for the IV is set to 1800 [DAC] without any multiplicity condition (cf. section \ref{IV_settings}).\\
Figure \ref{ISvsSUM_IV} shows another presentation of the data shown in Figure \ref{Multiplicity_vs_SUM}. For each IV group the fraction of events where the particular group was active is shown for different sum thresholds.

\noindent For a low sum threshold the average number of active groups is also low. The position of the corresponding active group over the IV is roughly uniform. For higher sum thresholds the fraction of active groups increases. For higher sum threshold values one can see that channels corresponding to the topological regions top (A), up (B) and lateral (C) have higher probabilities to be active than the ones located at the bottom of the IV. This can be explained by the fact that most muons enter the IV from above. The top group has a lower probability to be active than the other groups located at the top of the IV because of the connected PMTs of this group are located on an inner ring of the IV and are facing inwards, thus observing only a small volume of the IV. The same applies to the PMTs connected to the bottom group. The differences within a certain topological group can be explained by differences in the group threshold values and gains of the PMTs, which hadn't been optimized at the time this measurement was taken.\\

\subsection{Optimization of the AC Coupling at the Trigger Input Channels}

\begin{figure}[htb]
\centering
\subfloat[]{
\includegraphics[angle=0,width=0.45\textwidth]{{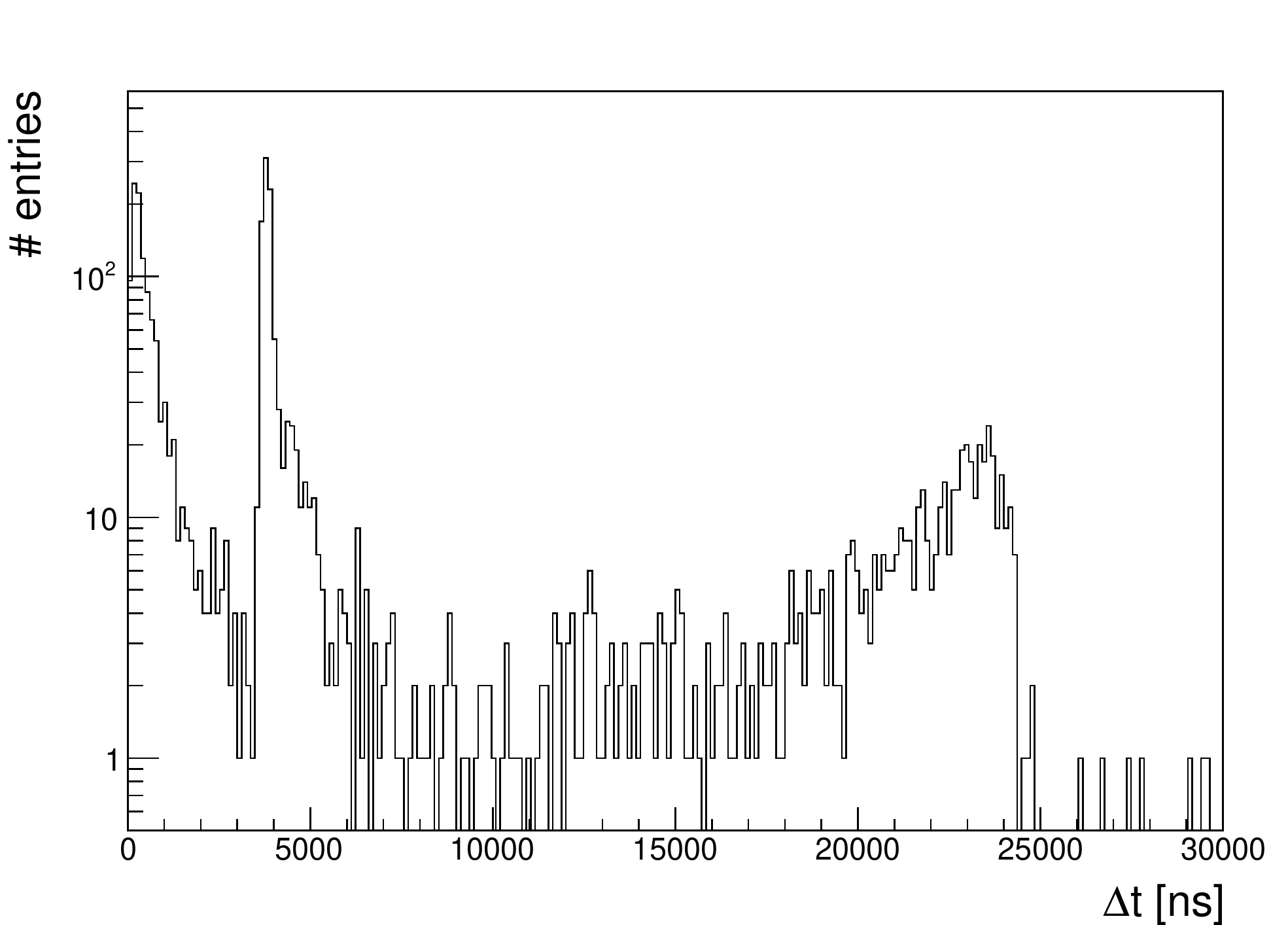}}
\label{DeltaT_v1}
}
\subfloat[]{
\includegraphics[angle=0,width=0.45\textwidth]{{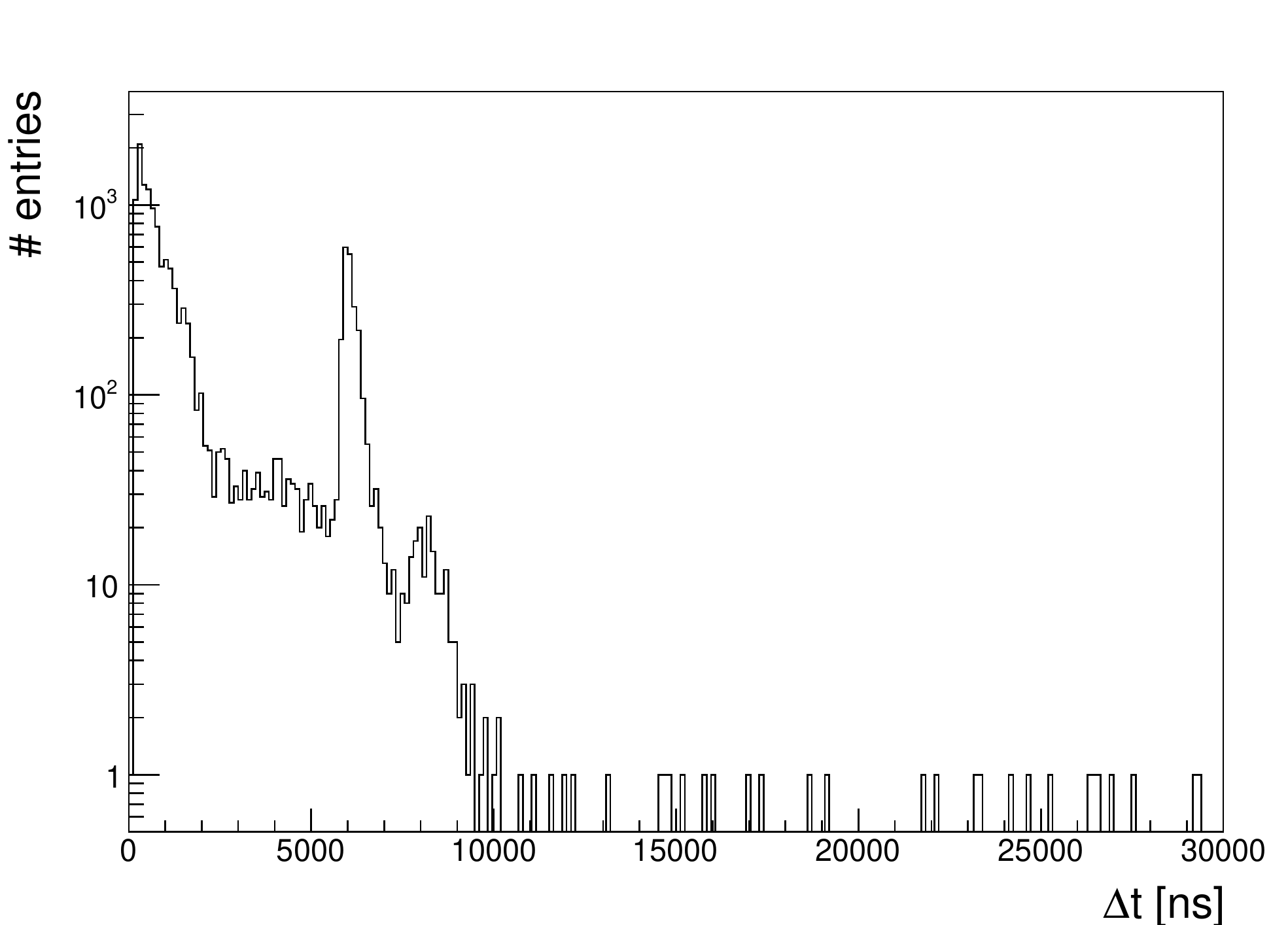}}
\label{DeltaT_v3}
}
\caption[]{(a): $\Delta$T spectrum of consecutive triggers, taken with an ID TB before the modification of the AC coupling. A positive overshoot of the group signals after muons cause an effective dead time between ~4.5 $\mu$s and ~24 $\mu$s due to the AC coupling at the TB inputs. (b): $\Delta$T spectrum of consecutive triggers, taken with an ID TB after the optimization of the AC coupling at the trigger inputs. The trigger is now sensitive as soon as the input signal returns to its baseline after the overshooting. The peaks at 6 and 8 $\mu$s match with the afterpulse distributions of the ID PMTs.\\}
\label{ACmod_v1}
\end{figure}

The AC coupling at the trigger input is realized by a RC-circuit. During the commissioning of the Double Chooz detector it turned out that a muon hitting the detector produces such a high light level that it strongly saturated the analogue electronics in front of the trigger. With the first version of the trigger system muons caused an effective deadtime due to combined effect on the PMTs, the FEE and the TBs. A muon signal produces a large positive overshoot at the end of the group signal. Due to the AC coupling at the TB inputs this leads to a release of a trigger (peak at approximately 4.5 $\mu$s in Figure \ref{DeltaT_v1}) when the stretcher signal returns to its baseline which has drifted towards positive voltages. The corresponding trigger signal stays active for approximately 20 $\mu$s which corresponds to the recovery time of the AC coupled baseline. During this time it is not possible to release a second trigger (cf. section \ref{trig1}) causing an effective deadtime of the system. After this time the threshold is still close to the group signal level such that even small fluctuations can cause a trigger. Those triggers cause the peak at 24 $\mu$s in Figure \ref{DeltaT_v1}.

\noindent SPICE simulations have shown \cite{ChuckConv}, that an increase of the coupling capacitor would improve the trigger response to these large overshoots. In Figure \ref{ACmod_v1} the improvement due to this modification is shown. With the modified TB version the positive overshoot does not cause additional triggers. The trigger becomes sensitive as soon as the group signal returns to its baseline. This can be seen in Figure \ref{DeltaT_v3} where the expected signals from correlated afterpulses of the PMTs in response to the initial high light level are triggered. The peak positions at 6 $\mu$s and 8 $\mu$s match with the afterpulse probability distribution of the ID PMTs \cite{APpaper}.\\
\noindent With the modification to the AC coupling at the TB inputs the trigger response shortly after high light level signals (e.g. caused by a muon) has been improved. Note, that for the neutrino analysis this is not an issue because a veto cut of 1 ms after each muon is applied \cite{DCFirstPub}, \cite{DC2ndPub}. With the improved AC coupling it becomes possible to trigger events even shortly after a muon (e.g. search for michel electrons).

\subsection{Trigger Dead Time Monitor Systems}
\label{DTM}

Due to the high energy deposited, muon events can cause the group signals to overshoot, making the trigger system less efficient at detecting lower energy events immediately after. In order to determine the impact of this inefficiency,  two dead time monitor systems were installed. The basic function of these dead-time monitors is to trigger the detector at a known, regular frequency, creating tagged random triggers with respect to physics triggers. By identifying any loss of such tagged triggers, we can quantify and track the dead time of the readout versus time.\\
Both triggers of the dead-time monitors run at 1 Hz, however they differ in their synchronization with the system clock. The first one is synchronous, created by the fixed rate trigger (cf. section \ref{trig1}) of the TMB. The second one is asynchronous, two NIM signals (with a fixed time difference of 2 $\mu$s) are produced by NIM gate generators once per second. When excluding muons and all following events for a duration of 1ms (offline veto time for neutrino searches), both dead time monitors show 100\% livetime; i.e. the trigger system does not introduce any dead-time for the neutrino analysis.

\section{Conclusions}

The trigger and timing system of the Double Chooz experiment was successfully installed and commissioned in 2011. It consists of three trigger boards (two for the inner detector and one for the IV), one trigger master board and several fan-outs. The trigger system is dead time less for the neutrino analysis. It continuously monitors the analogue signals of the detector and triggers the data acquisition based on a decision involving a logical AND of a threshold on the sum signal and a multiplicity condition representing the number of active groups of input channels. The trigger and timing system synchronizes all electronics and provides a time stamp, an event classification and an event number for each triggered event. The precision and stability of the common clock is well below the requirements of the experiment. However, the implementation of the custom GPS board would provide the possibility to participate in a global campaign of the detection of neutrinos from super nova neutrinos.\\
For the first data taking period of the Double Chooz experiment the physics threshold of the ID is determined to approximately 350 keV \cite{DCFirstPub, DC2ndPub}. The trigger efficiency is 100.0 \% above the analysis threshold of 0.7 MeV with a negligible uncertainty for the neutrino analysis.\\
The design of the trigger and timing system provides a high flexibility for the applied logic and settings, making it useful for experiments other than Double Chooz.

\section{Acknowledgments}

This work is supported by the DFG (Deutsche Forschungsgemeinschaft).\\
We thank the Double Chooz online/electronics group for the excellent cooperation. We thank C. Lane et al. of Drexel University for his help on the modification of the AC coupling of the TBs and for a FEE module they provided us for test measurements in our laboratory. Furthermore we thank A.Verdugo et al. of CIEMAT, Madrid for providing a splitter box for tests.\\
This paper is dedicated to our chief electronics engineer Franz Beissel and his family. Franz passed away during the development of this manuscript. The development of the trigger system has been a main focus of his recent work and we will deeply miss him and his expertise.

\end{document}